# A cluster mean approach for topology optimization of natural frequencies and bandgaps with simple/multiple eigenfrequencies

Shiyao Sun[1] and Kapil Khandelwal[2]

[1]Graduate Student, Dept. of Civil & Env. Engg. & Earth Sci., University of Notre Dame, Notre Dame, IN 46556, United States.

[2]Associate Professor, Dept. of Civil & Env. Engg. & Earth Sci., 156 Fitzpatrick Hall, University of Notre Dame, Notre Dame, IN 46556, United States, Email: kapil.khandelwal@nd.edu
ORCID: 0000-0002-5748-6019, (Corresponding Author)

*Preprint Submitted*

## Abstract

This study presents a novel approach utilizing cluster means to address the non-differentiability issue arising from multiple eigenvalues in eigenfrequency and bandgap optimization. By constructing symmetric functions of repeated eigenvalues – including cluster mean, $p$-norm and KS functions – the study confirms their differentiability when all repeated eigenvalues are included, i.e., clusters are complete. Numerical sensitivity analyses indicate that, under some symmetry conditions, multiple eigenvalues may also be differentiable w.r.t the symmetric design variables. Notably, regardless of enforced symmetry, the cluster mean approach guarantees differentiability of multiple eigenvalues, offering a reliable solution strategy in eigenfrequency topology optimization. Optimization schemes are proposed to maximize eigenfrequencies and bandgaps by integrating cluster means with the bound formulations. The efficacy of the proposed method is demonstrated through numerical examples on 2D and 3D solids and plate structures. All optimization results demonstrate smooth convergence under simple/multiple eigenvalues.

# 1 Introduction

Topology optimization of structures by tuning the natural frequencies for vibration control finds broad applications in engineering. In the design of structures and machines, it is desirable to optimize the natural frequencies to achieve increased capacity to withstand potential dynamic loads [1, 2]. In situations requiring vibration attenuation, a practical solution involves driving the natural frequencies of a structure away from external excitation frequencies to mitigate the risk of resonance [3]. Another application is to limit the natural frequencies of a system to remain outside a prescribed frequency range, i.e., creating frequency bandgaps [4]. Conversely, in applications where large-magnitude oscillations are desired, such as in sensing [5, 6], resonating actuators [7, 8], and energy harvesting [9, 10], engineers may tune the frequencies of devices towards the external excitation frequency to maximize the effect of such excitations. In applications where both the eigenfrequencies and eigenmodes are of engineering concern, topology optimization with eigenvalue and eigenvector constraints can be applied to achieve design goals [8, 11]. Natural frequencies – as intrinsic properties of a structure – thus closely correlate with the overall performance of the system. Consequently, in the design of engineered systems for vibration control there is a need for optimizing both the magnitude of natural frequencies and frequency bandgaps.

Optimization of natural frequencies falls into a broad category of problems that can be classified as eigenvalue optimization, as natural frequencies are eigenvalues of a generalized system [12]. Among other design methods, topology optimization stands out as an excellent tool for achieving this goal [2], as it provides considerable design flexibility. In topology optimization methods, gradient-based optimization algorithms are commonly employed as these methods have better performance than heuristic methods [13]. Moreover, in largescale topology optimization problems, the first-order algorithms are preferred, as second-order derivatives/sensitivities are expensive to compute [14]. Thus, for the success of gradient-based algorithms in eigenvalue optimization, the first-order sensitivities of the objective and constraint functions must be correctly computed. While the calculation of sensitivities of simple eigenvalues w.r.t design variables is straightforward, the same calculation cannot be applied to eigenvalues with multiplicities greater than one. This is because multiple/repeated eigenvalues are not always Fréchet differentiable and maybe only Gâteaux differentiable, i.e., only directional derivatives exist [15, 16]. Multiple eigenfrequencies can occur in complex structures with symmetries [16, 17] and are usually present in topology optimization problems where eigenvalues are optimized [18-20].

Two main categories of methods are adopted in literature to address the non-differentiability of repeated eigenvalues in eigenvalue optimization. In cases where the multiple eigenvalues are Gâteaux differentiable, directional derivatives are frequently used in topology optimization problems, among others [21-24]. The method of using directional derivatives in sensitivity analysis is now well-established [18, 19, 25-29]. In this method, to ensure that the multiple eigenvalues are directionally differentiable in the special directions of the perturbation in design variables, the optimization problem must be reformulated. To this end, the objective and constraint functions are expressed in terms of the *incremental* design variables when multiple eigenvalues are encountered during optimization. Moreover, in this new formulation, additional constraints need to be introduced. These extra constraints ensure that the *first-order* corrections of multiple eigenvalues are *linear* in the change in design variables. The overall optimization algorithm becomes more involved as it switches between different formulations – depending on the presence/absence of multiple eigenvalues. Interested readers are referred to Refs. [18, 19] for more details.

The second approach for addressing the non-differentiability arising from multiple eigenvalues entails constructing a symmetric function of eigenvalues. A symmetric function is such that its value is invariant under the permutation of its arguments. The main benefit of using symmetric functions is that a symmetric function of eigenvalues in the spectrum, regardless of their multiplicities, is differentiable [15, 30]. Symmetric functions such as the $p$-norm and Kreisselmeier-Steinhauser (KS) [31] can be used in eigenvalue topology optimization to resolve the non-differentiability of multiple eigenvalues. However, using symmetric functions is less explored compared to the directional derivative method. Past studies have utilized the $p$-norm function [32] and the KS function [32-34] to provide a differentiable approximation of the max/min functions. In Refs. [35-37], the $p$-norm function is utilized to resolve both the non-smoothness of the max/min function and the non-differentiability of multiple eigenvalues. Although effective in some problems, these aggregation functions provide only approximations of the extremum of a set of eigenvalues. Nonetheless, the concept of employing symmetric functions is desirable because of the lower algorithmic complexity, i.e., no reformulations or additional constraints are needed, as compared to the optimization algorithms based on directional derivatives. Building upon its inherent advantages, this work adopts the symmetric function approach to address the non-differentiability issue associated with multiple eigenvalues effectively.

The main contribution of this work is to introduce a computationally efficient eigenfrequency topology optimization framework that utilizes a cluster mean approach together with bound formulations, which ensure the differentiability of eigenvalues of arbitrary multiplicity. While the bound formulation systematically resolves the non-smoothness issue inherent in the max-min problems in eigenfrequency optimization [38, 39], the proposed cluster mean approach effectively tackles the non-differentiability associated with repeated eigenvalues. In the proposed cluster mean approach, prior to constructing a mean function that is symmetric polynomial, the computed eigenvalues first undergo a clustering operation to form distinct eigen-clusters. This clustering assigns multiple (repeated) eigenvalues to the same cluster. In this way, instead of optimizing individual eigenfrequencies, the mean functions of the eigen-clusters are optimized. Using these differentiable mean functions, new formulations based on bound variables are proposed to maximize any specified order of eigenfrequency or bandgap. Each cluster mean can be more accurately controlled in the proposed formulation, distinguishing it from the aggregation function method, which merely approximates the extrema of eigenfrequencies. In addition, numerical studies are carried out using the central difference method to verify the differentiability of symmetric functions based on eigenvalue clusters. Based on the numerical results, further clarifications are made on the construction and use of symmetric functions. Specifically, it is shown that symmetric functions such as mean, $p$-norm, and KS functions are only differentiable if all repeated eigenvalues are included in the considered function. This is an important result as these functions are frequently employed in topology optimization but *without* clustering. Furthermore, it is numerically shown that the multiple eigenvalues might even be differentiable depending on the underlying symmetry. Finally, the efficacy of proposed bound formulations based on the cluster mean approach in optimizing eigenfrequency is demonstrated through numerical experiments. These involve optimizing the eigenfrequencies and bandgaps of 2D and 3D solids and multi-material Reissner-Mindlin plate structures.

The rest of the paper is organized as follows. Section 2 provides the theoretical background for the finite element eigen-analysis (Section 2.1), material interpolation schemes (Section 2.2), topology optimization formulations (Section 2.3), and sensitivity verification studies considering two different enforced symmetries (Section 2.4). Section 3 investigates the capability of the cluster mean approach and the bound formulation in eigenfrequency topology optimization through a range of numerical examples. The conclusions of this work can be found in Section 4.

## 2 Problem Formulation

### 2.1 Finite Element Eigen-analysis

Finite element analysis (FEA) is used to evaluate the structural eigenfrequencies at each iteration step of topology optimization. In FEA, the system domain $\Omega$ is discretized by a mesh consisting of finite elements. The discretized (global) generalized eigenvalue problem is given by [12]

$$\boldsymbol{K}\boldsymbol{\phi}_q = \lambda_q \boldsymbol{M}\boldsymbol{\phi}_q, \quad q = 1, 2, \ldots, n_f \tag{1}$$

where $\boldsymbol{K}$ and $\boldsymbol{M}$ are the global stiffness and mass matrices considering the free degrees of freedom, respectively; $\lambda_q \geq 0$ and $\boldsymbol{\phi}_q$ are the eigenvalues and eigenvectors of the generalized eigensystem in Eq. (1), respectively; and $n_f$ are the total number of (free) degrees of freedom in the system. The eigenvalues $\lambda_q$ are listed in increasing order, i.e., $\lambda_1 \leq \lambda_2 \leq \cdots \leq \lambda_{n_f}$, and $\omega_q \overset{\text{def}}{=} \sqrt{\lambda_q}$ is the $q^{th}$ eigenfrequency of the system. Note that $\boldsymbol{K}$ is a symmetric positive definite matrix while $\boldsymbol{M}$ is assumed to be symmetric positive semidefinite. Moreover, the eigenvectors are $\boldsymbol{M}$-orthonormalized, i.e., $\boldsymbol{\phi}_a^T \boldsymbol{M} \boldsymbol{\phi}_b = \delta_{ab}$ and $\boldsymbol{\phi}_a^T \boldsymbol{K} \boldsymbol{\phi}_b = \lambda_b \delta_{ab}$, where $\delta_{ab} = \begin{cases} 1 & a = b \\ 0 & a \neq b \end{cases}$ is the Kronecker symbol. Stiffness matrix $\boldsymbol{K}$ and mass matrix $\boldsymbol{M}$ are assembled using corresponding element stiffness ($\boldsymbol{K}^e$) and mass ($\boldsymbol{M}^e$) matrices.

  a) 2D and 3D Solids

Element stiffness matrix ($\boldsymbol{K}^e$):

$$\boldsymbol{K}^e = \int_{\Omega^e} [\boldsymbol{B}^e]^T [\mathbb{C}^e][\boldsymbol{B}^e] \, d\Omega \tag{2}$$

where $\boldsymbol{B}^e$ is the shape function derivative matrix for 4-node quadrilateral (2D case) or 8-node brick (3D case) elements [12] and $n_{ele}$ is the total number of finite elements in the domain. The constitutive tensor $\mathbb{C}$ of the underlying linear isotropic material is given as $\mathbb{C}^e_{abcd} = \lambda \delta_{ab} \delta_{cd} + \mu(\delta_{ac}\delta_{bd} + \delta_{ad}\delta_{bc})$ where the Lamé parameters $\lambda$ and $\mu$, are related to the element Young's modulus $E^e$ and Poisson's ratio $\nu$ by $\lambda = \frac{E^e \nu}{(1+\nu)(1-2\nu)}$ and $\mu = \frac{E^e}{2(1+\nu)}$. The element Young's modulus $E^e$ is dependent on design variables while the Poisson's ratio $\nu$ is fixed.

Element mass matrix ($\boldsymbol{M}^e$):

$$\boldsymbol{M}^e = \int_{\Omega^e} \rho_m^e [\boldsymbol{N}^e]^T [\boldsymbol{N}^e] \, d\Omega \tag{3}$$

where $\rho_m^e$ is the element mass density which depends on design variables, and $\boldsymbol{N}^e$ is the appropriate shape function matrix for 2D or 3D elements [12].

b) Reissner-Mindlin Plate Element

For eigenfrequency optimization of plate structures, the quadrilateral Reissner-Mindlin plate element proposed by Gruttmann & Wagner [40] is employed. This element does not exhibit shear locking, has the correct rank, and is shown to be effective for both thick and thin plates [40]. The element stiffness matrix consists of two components: one obtained by one-point integration and the other given by a stabilization matrix.

Plate element stiffness matrix ($\boldsymbol{K}^e$):

$$\boldsymbol{K}^e = \boldsymbol{K}_0^e + \boldsymbol{K}_{\text{stab}}^e \tag{4}$$

where $\boldsymbol{K}_0^e$ and $\boldsymbol{K}_{\text{stab}}^e$ are the one-point integrated stiffness and the stabilization matrices, respectively. These matrices depend on the effective constitutive matrix $\boldsymbol{C}_P^e$ for plate, consisting of bending ($\boldsymbol{C}_b^e$) and shear ($\boldsymbol{C}_s^e$) components, and is given by

$$\boldsymbol{C}_P^e = \begin{bmatrix} \boldsymbol{C}_b^e & \boldsymbol{0} \\ \boldsymbol{0} & \boldsymbol{C}_s^e \end{bmatrix}$$

$$\boldsymbol{C}_b^e = \frac{E^e h^3}{12(1-\nu^2)} \begin{bmatrix} 1 & \nu & 0 \\ \nu & 1 & 0 \\ 0 & 0 & \frac{1-\nu}{2} \end{bmatrix} \text{ and } \boldsymbol{C}_s^e = \frac{\kappa_s E^e h}{2(1+\nu)} \begin{bmatrix} 1 & 0 \\ 0 & 1 \end{bmatrix} \tag{5}$$

where $h$ is the plate thickness and the shear correction factor $\kappa_s = \frac{5}{6}$. The element Young's modulus $E^e$ is dependent on the design variable, while Poisson's ratio $\nu$ is fixed. More details concerning the Reissner-Mindlin plate element including the formulations of $K_0^e$ and $K_{stab}^e$ can be found in [40].

Plate element mass matrix ($M^e$):

$$M^e = \int_{\Omega^e} \rho_m^e [N^e]^T [c_m][N^e] d\Omega \quad \text{with} \quad [c_m] = \begin{bmatrix} h & 0 & 0 \\ 0 & \frac{h^3}{12} & 0 \\ 0 & 0 & \frac{h^3}{12} \end{bmatrix} \tag{6}$$

where $\rho_m^e$ is the element mass density which depends on design variables, and $N^e$ is the shape function matrix for the plate element.

### 2.2 Stiffness and Mass Interpolation

In the considered density-based topology optimization framework, a topology is parameterized by elementwise density variables that represent the volume fraction of the material phases available for the considered design. The specific interpolation scheme depends on the number of material phases considered.

  a) Solid-void and Bi-material interpolation

In the solid-void case, the topology is parameterized by an elementwise-constant material phase density field represented by $\{\rho_e\}_{e=1}^{n_{ele}}$, where $\rho_e = 1$ and $\rho_e = 0$ represents the presence and absence of material in the $e^{th}$ element, respectively. The density variables $\rho \in \mathbb{R}^{n_{ele}}$ that are used for material interpolation are obtained from design variables $x \in \mathbb{R}^{n_{ele}}$ ($x_{min} \leq x \leq 1$) via a density filter [41-43], i.e., $x \xmapsto{W} \rho$, which is used to address mesh dependency and checkerboarding issues. This (linear) filter operation is given by

$$\rho = Wx \quad \text{with} \quad W_{pq} = \frac{w_{pq} v_q}{\sum_{q=1}^{n_{ele}} w_{pq} v_q} \quad \text{and} \quad w_{pq} = \max\left(r_{min} - \|X_p - X_q\|_2, 0\right) \tag{7}$$

where $v_q$ is the volume of the $q^{th}$ element, $r_{min}$ is the filter radius and $X_p$ is the centroidal coordinates of the $p^{th}$ element. To facilitate gradient-based optimization, the element density variable $\rho_e$ is relaxed such that $\rho_e \in [\rho_L, 1.0]$, where $\rho_L$ is a small number introduced to prevent the singularity of the stiffness matrix. With the density filter Eq. (7), $\rho_L$ is dependent on $x_{\min}$ which is the minimum design variable and is set to $10^{-4}$ in all solid-void phase problems.

Another issue in eigenfrequency optimization using standard Solid Isotropic Material with Penalization (SIMP) interpolation is the appearance of pseudo eigenmodes in low-density regions [44]. To mitigate the effect of these pseudo eigenmodes on the optimization process, the idea of modifying the interpolated stiffness and mass matrices in low-density elements is frequently used. Many studies have demonstrated the effectiveness of the modified interpolation approach in removing the pseudo eigenmodes [7, 18, 44]. This work follows the modified SIMP interpolation scheme proposed in [18] to address this issue. To this end, the solid-void phase Young's modulus ($E^e$) and mass density ($\rho_m^e$) of a finite element are interpolated in terms of the element density variable ($\rho_e$) using the modified SIMP method as

$$E^e(\rho_e) = \rho_e^{p_1} E_s \quad \rho_L \leq \rho_e \leq 1$$

$$\rho_m^e(\rho_e) = \begin{cases} (c_1 \rho_e^{p_2} + c_2 \rho_e^{p_2+1})\rho_s & \rho_L \leq \rho_e \leq \rho_T \\ \rho_e \rho_s & \rho_T \leq \rho_e \leq 1 \end{cases} \tag{8}$$

where $E_s$ and $\rho_s$ represent Young's modulus and mass density of the solid material phase, respectively; $p_1 \geq 1$ is an implicit penalty parameter to penalize intermediate densities. With this modified SIMP approach, when an element density $\rho_e$ is smaller than the threshold $\rho_T$, an extra penalization is introduced to the mass density interpolation. The main idea is to impose a higher penalty on the mass density than on the Young's modulus in low-density elements. This approach aims to push low-density eigenmodes to higher frequencies, and so they are not included in the optimization. In this study, the penalty parameters of Young's modulus, $p_1$, and mass density, $p_2$, are chosen as 3 and 6, respectively. The coefficients $c_1$ and $c_2$ are in Eq. (8) are calculated to enforce $C^1$-continuity of the density function $\rho_m^e(\rho_e)$ at $\rho_e = \rho_T$. More details on this modified SIMP interpolation scheme can be found in [18].

The above material interpolation scheme for the solid-void phase is extended to the bi-material problems (without void phase) as

$$E^e(\rho_e) = \rho_e^p E_{s_1} + (1 - \rho_e^p) E_{s_2} \quad 0 \leq \rho_e \leq 1$$

$$\rho_m^e(\rho_e) = \rho_e \rho_{s_1} + (1 - \rho_e) \rho_{s_2} \quad 0 \leq \rho_e \leq 1 \qquad (9)$$

where $E_{s_1}$ and $E_{s_2}$ represents the Young's moduli of the solid material phases 1 and 2, respectively. Similarly, $\rho_{s_1}$ and $\rho_{s_2}$ denote the mass densities of solid material phases 1 and 2, respectively. In this case, the penalty parameter, $p$, is taken as 3. The density variable $\rho_e$, in this case, represents the volume fraction of material phase-1, while $(1 - \rho_e)$ is the volume fraction of material phase-2. It is noted that in the bi-material phase, without voids, the SIMP interpolations for Young's modulus and mass density do not require extra modifications and $x_{\min} = 0.0$ is used.

b) Multi-material interpolation

Topology optimization studies on plates with bi-material + void phase and tri-material + void phase are also considered. Mixture rules for multi-materials are used to construct these schemes [45, 46]. Interpolations for Young's modulus ($E^e$) and mass density ($\rho_m^e$) for these cases with the modified SIMP method are given by:

(1) Bi-material + void phase

Young's modulus ($E^e$):
$$E^e(\rho_1, \rho_2) = \rho_1^{p_1} E_{12}^e \quad \rho_L \leq \rho_1 \leq 1$$

$$E_{12}^e(\rho_2) = \rho_2^{p_1} E_{s_1} + (1 - \rho_2^{p_1}) E_{s_2} \quad 0 \leq \rho_2 \leq 1 \qquad (10)$$

Mass density ($\rho_m^e$):

$$\rho_m^e(\rho_1, \rho_2) = \begin{cases} (c_1\rho_1^{p_2} + c_2\rho_1^{p_2+1})\rho_{12}^e & \rho_L \leq \rho_1 \leq \rho_T \\ \rho_1\rho_{12}^e & \rho_T \leq \rho_1 \leq 1 \end{cases}$$

$$\rho_{12}^e(\rho_2) = \rho_2\rho_{s_1} + (1-\rho_2)\rho_{s_2} \quad 0 \leq \rho_2 \leq 1$$

(11)

Note that the subscripts $\blacksquare_e$ of elementwise density variables $\rho_1$ and $\rho_2$ are omitted to simplify notations and it is understood that these quantities correspond to the $e^{th}$ element. In this case the design variables $x = [x_1, x_2]$ such that $x_1, x_2 \in \mathbb{R}^{n_{ele}}$, with $x_{1,\min} \leq x_1^e \leq 1$, $0 \leq x_2^e \leq 1$, and $\rho_l = Wx_l$ with $l = 1$ and 2, i.e., the filter is applied to individual material phases. Here, $\rho_1$ controls the element volume fraction of the solid/void phases, while $\rho_2$ controls the distribution of two solid material phases. Accordingly, $x_{1,\min} = 1.0E\text{-}4$ is used, while $x_{2,\min} = 0$.

(2) Tri-material + void phase

Young's modulus ($E^e$):

$$E^e(\rho_1, \rho_2, \rho_3) = \rho_1^{p_1} E_{123}^e \quad \rho_L \leq \rho_1 \leq 1$$

$$E_{123}^e(\rho_2, \rho_3) = \rho_2^{p_1} E_{12}^e + (1 - \rho_2^{p_1})E_{s_3} \quad 0 \leq \rho_2 \leq 1$$

(12)

$$E_{12}^e(\rho_3) = \rho_3^{p_1} E_{s_1} + (1 - \rho_3^{p_1})E_{s_2} \quad 0 \leq \rho_3 \leq 1$$

Mass density ($\rho_m^e$):

$$\rho_m^e(\rho_1, \rho_2, \rho_3) = \begin{cases} (c_1\rho_1^{p_2} + c_2\rho_1^{p_2+1})\rho_{123}^e & \rho_L \leq \rho_1 \leq \rho_T \\ \rho_1\rho_{123}^e & \rho_T \leq \rho_1 \leq 1 \end{cases}$$

$$\rho_{123}^e(\rho_2, \rho_3) = \rho_2\rho_{12}^e + (1-\rho_2)\rho_{s_3} \quad 0 \leq \rho_2 \leq 1$$

(13)

$$\rho_{12}^e(\rho_3) = \rho_3\rho_{s_1} + (1-\rho_3)\rho_{s_2} \quad 0 \leq \rho_3 \leq 1$$

Here, $E_{s_1}$, $E_{s_2}$, and $E_{s_3}$ represent Young's moduli of solid material phases 1, 2, and 3, respectively. Likewise, $\rho_{s_1}$, $\rho_{s_2}$, and $\rho_{s_3}$ represent mass densities of solid material phases 1, 2, and 3, respectively. The penalty parameters of Young's modulus, $p_1$, and mass density, $p_2$, are chosen as 3 and 6, respectively. In this case, the design variables $x = [x_1, x_2, x_3]$ such that $x_1, x_2, x_3 \in \mathbb{R}^{n_{ele}}$, with $x_{1,\min} \leq x_1^e \leq 1$, $0 \leq x_2^e, x_3^e \leq 1$, and $\rho_l = Wx_l$ with $l = 1, 2$, and 3, i.e., the filter is again applied to individual material phases. Similarly, $\rho_1$ controls the element volume fraction of the solid/void phases, while $\rho_2$ and $\rho_3$ control the distribution of three solid material phases. Accordingly, $x_{1,\min} = 1.0\text{E-}4$ is used, while $x_{2,\min} = x_{3,\min} = 0$.

## 2.3 Topology Optimization Formulations

Different optimization formulations are introduced using the bound formulation where the objective is to maximize a target eigenfrequency or a target bandgap. These formulations are constructed utilizing a cluster mean approach to ensure differentiability. The cluster mean approach was first introduced by [20] to address the non-differentiability of repeated eigenvalues in buckling-constrained topology optimization. This approach is further extended to eigenfrequency optimization formulations in the present study. To this end, the eigenvalues in Eq. (1) are first clustered according to their multiplicity, i.e., $\underbrace{\lambda_1 = \cdots = \lambda_{N_1}}_{\Lambda_1} < \underbrace{\lambda_{N_1+1} = \cdots = \lambda_{N_1+N_2}}_{\Lambda_2} < \cdots < \underbrace{\lambda_{N_1+\cdots+N_{n_c-1}+1} = \cdots = \lambda_{N_1+\cdots+N_{n_c}}}_{\Lambda_{n_c}}$ with $n_T := N_1 + \cdots + N_{n_c}$, $N_q \geq 1$. The number of such required clusters, $n_c$, depends on the considered problem formulation. In general, $n_T \ll n_f$, and therefore, only a limited number of eigenvalues are needed. Once the clustering is established, the mean value $\overline{\Lambda}_q$ of the $q^{th}$ cluster, which is a symmetric function of the eigenvalues, is calculated as

$$\overline{\Lambda}_q = \frac{1}{N_q} \sum_{\lambda_k \in \Lambda_q} \lambda_k \tag{14}$$

The main benefit of the cluster mean ($\overline{\Lambda}_q$) in Eq. (14) is that this symmetric function is differentiable even if the individual repeated eigenvalues in the cluster are not differentiable. This differentiable object is now used to construct appropriate bound formulations that remain differentiable. In practice, the clustering is done numerically, and an eigenvalue is assigned to the same cluster if the relative difference between the minimum eigenvalue in the cluster and the considered eigenvalue is less than a prescribed tolerance $\epsilon_{\text{tol}}$. For instance, for eigen-cluster $\Lambda_1$, $\frac{|\lambda_r - \lambda_1|}{|\lambda_1|} \leq \epsilon_{\text{tol}}$ where $r = 2, \ldots, N_1$. A relative tolerance of $\epsilon_{\text{tol}} = 1.0\text{E-}8$ is used for clustering eigenvalues and the effect of this tolerance parameter is illustrated via numerical example.

*2.3.1.1 Maximization of target eigen-cluster mean*

The bound formulation of the solid-void phase for maximizing the $n^{th}$ eigen-cluster mean is constructed in terms of the mean of the $m$ eigenclusters $\overline{\Lambda}_n < \overline{\Lambda}_{n+1} < \cdots < \overline{\Lambda}_{n+m-1}$ as follows

$$\max_{\mathbf{y}} f_0(\mathbf{y}) = \beta$$

$$\text{s.t. } f_1(\mathbf{y}) = \frac{\beta}{\overline{\Lambda}_n} - 1 \leq 0$$

$$f_2(\mathbf{y}) = \frac{\beta}{\overline{\Lambda}_{n+1}} - 1 \leq 0$$

$$\vdots$$

$$f_m(\mathbf{y}) = \frac{\beta}{\overline{\Lambda}_{n+m-1}} - 1 \leq 0$$

$$f_{m+1}(\mathbf{y}) = \frac{1}{V}\left(\sum_{e=1}^{n_{ele}} \rho_e v_e\right) - V_f \leq 0$$

$$\beta > 0, \mathbf{x}_{\min} \leq \mathbf{x} \leq \mathbf{1}$$

(15)

where $\mathbf{y} = \{\beta, \mathbf{x}\}$. In the above formulation, $\beta$ is the bound variable; $f_{m+1}$ is the volume constraint for the solid-void case, where $V$ is the total volume and $V_f$ is the allowed volume fraction threshold. All entries in $\mathbf{x}_{\min}$ are set as $\mathbf{x}_{\min} = 1.0\text{E-}4$. Constraints $f_1$ to $f_m$ involve the target $n^{th}$ eigen-cluster $\Lambda_n$, and together with the extra $(m-1)$ clusters $\Lambda_{n+1}, \ldots, \Lambda_{n+m-1}$ which have a greater mean value than the target cluster $\Lambda_n$. With this bound formulation, when $\beta$ is maximized, the constraints $f_1$ to $f_m$ on eigen-clusters ensure that the $n^{th}$ eigenfrequency cluster is also maximized. In this study, $m$ is set to be 10, i.e., a total of 10 eigenvalue cluster constraints are included in all numerical examples.

In the bi-material + void phase case, the two volume constraints $f_{m+1}$ and $f_{m+2}$ are enforced as

$$f_{m+1}(\mathbf{y}) = \frac{1}{V}\left(\sum_{e=1}^{n_{ele}} \rho_1^e v_e\right) - V_1^f \leq 0$$

$$f_{m+2}(\mathbf{y}) = \frac{1}{V}\left(\sum_{e=1}^{n_{ele}} \rho_1^e \rho_2^e v_e\right) - V_2^f \leq 0 \tag{16}$$

$$\beta > 0,\ \mathbf{x}_{1,\min} \leq \mathbf{x}_1 \leq 1,\ 0 \leq \mathbf{x}_2 \leq 1$$

where $\mathbf{y} = \{\beta, \mathbf{x} = [\mathbf{x}_1, \mathbf{x}_2]\}$. Here, $V_1^f$ and $V_2^f$ are the threshold values for the volume constraints for the bi-material + void phase case, where $V_1^f$ limits the ratio between the volume with materials and the total volume, and $V_2^f$ limits the ratio between the volume with material 1 and the total volume $V$. All entries in $\mathbf{x}_{1,\min}$ are set as $x_{1,\min} = 1.0\text{E-}4$.

In the tri-material + void phase case, three volume constraints $f_{m+1}$ $f_{m+2}$ and $f_{m+3}$ are incorporated as

$$f_{m+1}(\mathbf{y}) = \frac{1}{V}\left(\sum_{e=1}^{n_{ele}} \rho_1^e v_e\right) - V_1^f \leq 0$$

$$f_{m+2}(\mathbf{y}) = \frac{1}{V}\left(\sum_{e=1}^{n_{ele}} \rho_1^e \rho_2^e v_e\right) - V_2^f \leq 0 \tag{17}$$

$$f_{m+3}(\mathbf{y}) = \frac{1}{V}\left(\sum_{e=1}^{n_{ele}} \rho_1^e \rho_2^e \rho_3^e v_e\right) - V_3^f \leq 0$$

$$\beta > 0,\ \mathbf{x}_{1,\min} \leq \mathbf{x}_1 \leq 1,\ 0 \leq \mathbf{x}_2 \leq 1,\ 0 \leq \mathbf{x}_3 \leq 1$$

where $\mathbf{y} = \{\beta, \mathbf{x} = [\mathbf{x}_1, \mathbf{x}_2, \mathbf{x}_3]\}$. Here, $V_1^f$, $V_2^f$ and $V_3^f$ are the threshold values for the volume constraints for the tri-material + void phase case, where $V_1^f$ limits the ratio between the volume with materials and the total volume, $V_2^f$ limits the ratio of the volume with material 1 and 2 to the total volume, and $V_3^f$ limits the ratio of the volume with material 1 to the total volume. For the tri-material + void phase case, all entries in $x_{1,\min}$ are set as $x_{1,\min} = 1.0\text{E-}4$.

*2.3.1.2 Maximization of target bandgap*

For bandgap optimization, two independent bound variables, $\beta_1$ and $\beta_2$, are introduced. The difference between the two bound variables $\beta_2 - \beta_1$ is the bandgap that is to be maximized. To this end, the bound formulation for maximizing the bandgap between $n^{th}$ and $(n+1)^{th}$ eigenfrequency is constructed in terms of mean eigen-clusters $\bar{\Lambda}_1 < \bar{\Lambda}_2 < \cdots < \bar{\Lambda}_n < \bar{\Lambda}_{n+1} < \cdots < \bar{\Lambda}_{n+m}$ and the bound variables, $\beta_1$ and $\beta_2$, as follows

$$\max_{y} f_0(y) = \beta_2 - \beta_1$$

$$\text{s.t. } \left.\begin{aligned} f_1(y) &= 1 - \frac{\beta_1}{\bar{\Lambda}_1} \leq 0 \\ &\vdots \\ f_n(y) &= 1 - \frac{\beta_1}{\bar{\Lambda}_n} \leq 0 \end{aligned}\right\} \text{clusters to be minimized}$$

$$\left.\begin{aligned} f_{n+1}(y) &= \frac{\beta_2}{\bar{\Lambda}_{n+1}} - 1 \leq 0 \\ &\vdots \\ f_{n+m}(y) &= \frac{\beta_2}{\bar{\Lambda}_{n+m}} - 1 \leq 0 \end{aligned}\right\} \text{clusters to be maximized} \qquad (18)$$

$$f_{n+m+1}(y) = \frac{1}{V}\left(\sum_{e=1}^{n_{ele}} \rho_e v_e\right) - V_f \leq 0$$

$$\beta_1, \beta_2 > 0, \; x_{\min} \leq x \leq 1$$

where $y = \{\beta_1, \beta_2, x\}$. Here, $f_{n+m+1}$ is the volume constraint for the solid-void case and all entries in $x_{\min}$ are set as $x_{\min} = 1.0\text{E-}4$. Constraints $f_1$ to $f_n$ imply that the first $n$ clusters need to be smaller than $\beta_1$. Similarly, constraints $f_{n+1}$ to $f_{n+m}$ require that the remaining $m$ clusters be greater than $\beta_2$. Thus, using this bound formulation when the difference $(\beta_2 - \beta_1)$ is maximized, the constraints $f_1$ to $f_{n+m}$ ensure that the target bandgap is also maximized. Moreover, in the bandgap problems $m$ is set to 10. Thus, there are a total number of $(n + 10)$ eigenvalue cluster constraints in the bandgap problem formulation. For the multi-material + void phase problem, the volume constraints are formulated as in Eq. (16) for the bi-material + void case and Eq. (17) for the tri-material + void case. The Method of Moving Asymptotes (MMA) is used as an optimizer [47], and the MMA move limit is set to 0.05. Except for the move limit, all other parameters in MMA are configured with default values.

*2.4 Differentiability and Sensitivity Analyses*

The primary goal of this work is to employ the cluster mean approach to address the non-differentiability problem associated with repeated eigenvalues in eigenfrequency optimization. First, consider the simple eigenvalue case: let $\lambda_q$ be a simple eigenvalue such that $\boldsymbol{K}\boldsymbol{\phi}_q = \lambda_q \boldsymbol{M}\boldsymbol{\phi}_q$, then the sensitivity of $\lambda_q$ w.r.t the $r^{th}$ density variable, $\rho_r$, is given by

$$\frac{d\lambda_q}{d\rho_r} = \boldsymbol{\phi}_q^T \left( \frac{d\boldsymbol{K}}{d\rho_r} - \lambda_q \frac{d\boldsymbol{M}}{d\rho_r} \right) \boldsymbol{\phi}_q \qquad (19)$$

Eq. (19) can be straightforwardly derived using direct differentiation [1]. However, Eq. (19) is not valid when the multiplicity of the eigenvalue is greater than 1 because the repeated eigenvalue might not be Fréchet differentiable. However, a symmetric function of the eigenvalues is still differentiable provided that all the repeated eigenvalues are included in the construction of such a symmetric function [15, 30]. It is emphasized that all the repeated eigenvalues must be included, or else the symmetric function will still be non-differentiable. This critical point is somewhat overlooked in past studies, and symmetric functions such as $p$-norm and KS functions are employed without adequate consideration of such multiplicities. This issue will be further demonstrated through numerical examples in the following section. In a similar vein, the mean-cluster function $\overline{\Lambda}_q$ in Eq. (14) is a differentiable function, i.e.,

$$\frac{d\bar{\Lambda}_q}{d\rho_r} = \frac{1}{N_q}\left(\sum_{\lambda_k \in \Lambda_q}\left[\boldsymbol{\phi}_k^T\left(\frac{d\boldsymbol{K}}{d\rho_r} - \lambda_k\frac{d\boldsymbol{M}}{d\rho_r}\right)\boldsymbol{\phi}_k\right]\right) \quad (20)$$

is well-defined, where $(\lambda_k, \boldsymbol{\phi}_k)$ are the eigenpairs in the cluster $\Lambda_q$. With this caveat, the sensitivity of the objective and constraint functions in Eq. (15) and Eq. (18) can be straightforwardly computed.

Another issue that is not yet well-understood is that the repeated eigenvalues may even be Fréchet differentiable – depending on the underlying *symmetry* of the considered problem [48]. It is not difficult to construct analytical examples where repeated eigenvalues are Fréchet differentiable, e.g., $\boldsymbol{K}(\boldsymbol{x}) = \begin{bmatrix} 1 + x_1 + x_1^2 & x_2^3 \\ x_2^3 & 1 + x_1 + x_1^2 \end{bmatrix}$ has eigenvalues $\{\lambda_1(\boldsymbol{x}) = 1 + x_1 + x_1^2 - x_2^3,\ \lambda_2(\boldsymbol{x}) = 1 + x_1 + x_1^2 + x_2^3\}$. When $x_2 = 0$, the two eigenvalues $\lambda_1(\boldsymbol{x}) = \lambda_2(\boldsymbol{x}) = 1 + x_1 + x_1^2$ become repeated, and the multiple eigenvalues remain differentiable everywhere. In the following section, the differentiability of symmetric functions of eigenvalues is studied using numerical examples. To illustrate the effect of problem symmetry, two cases with low and high underlying symmetries are considered. Both cases admit repeated eigenvalues. Using these test cases, the differentiability of symmetric functions, i.e., mean, $p$-norm, and KS functions, is verified. The sensitivity analysis compares the analytical sensitivities with those calculated by the central difference method (CDM). A perturbation value of $\Delta h = 1.0\text{E-}8$ is used to obtain numerical sensitivities via CDM.

### 2.4.1 Symmetry Type-1: ½-symmetry

For sensitivity verification, the square block design domain shown in Figure 1 is discretized into 20×20 4-node quadrilateral elements. The boundary condition includes fixed supports at each corner of the 2D square block. Notably, the design space admits ½-symmetry and such symmetry is enforced. Sensitivity analysis is carried out on a design obtained after 10 optimization iterations, where the objective is to maximize the first eigen-cluster mean $\bar{\Lambda}_1$ with the density filter radius of 0.6, and the volume fraction is restricted to 0.5. A total of 10 eigen-clusters are included in the constraints in optimization formulation in Eq. (15). The intermediate design after 10 iterations and convergence history are shown in Figure 2. The eigen-cluster history shown in Figure 2(right) indicates that the eigen-clusters $\Lambda_1$, $\Lambda_6$ and $\Lambda_8$ each has 2 repeated eigenvalues. The detailed information of the eigen-clusters at iteration step 10 is summarized in Table 1. The sensitivity analysis is first carried out on the individual multiple eigenvalues and then on 3 symmetric functions: mean, $p$-norm, and KS functions. Moreover, the sensitivity analysis compares the sensitivities computed by the analytical method (Eq. (19)) and CDM w.r.t all the design variables, i.e., ignoring underlying symmetry, and with w.r.t only symmetric design variables.

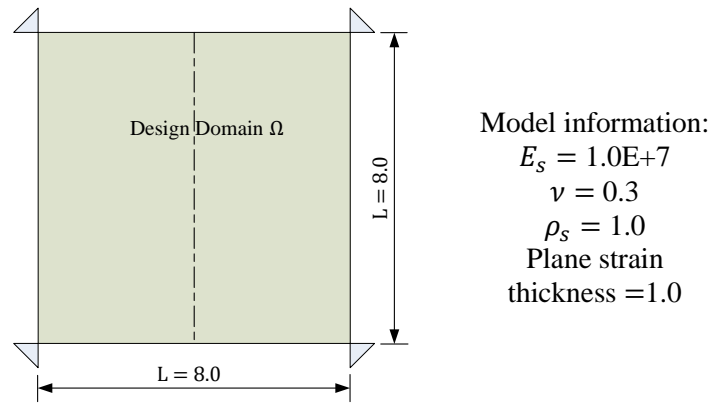

Figure 1. Square block design domain – dashed line indicates the axis of symmetry (½-symmetry).

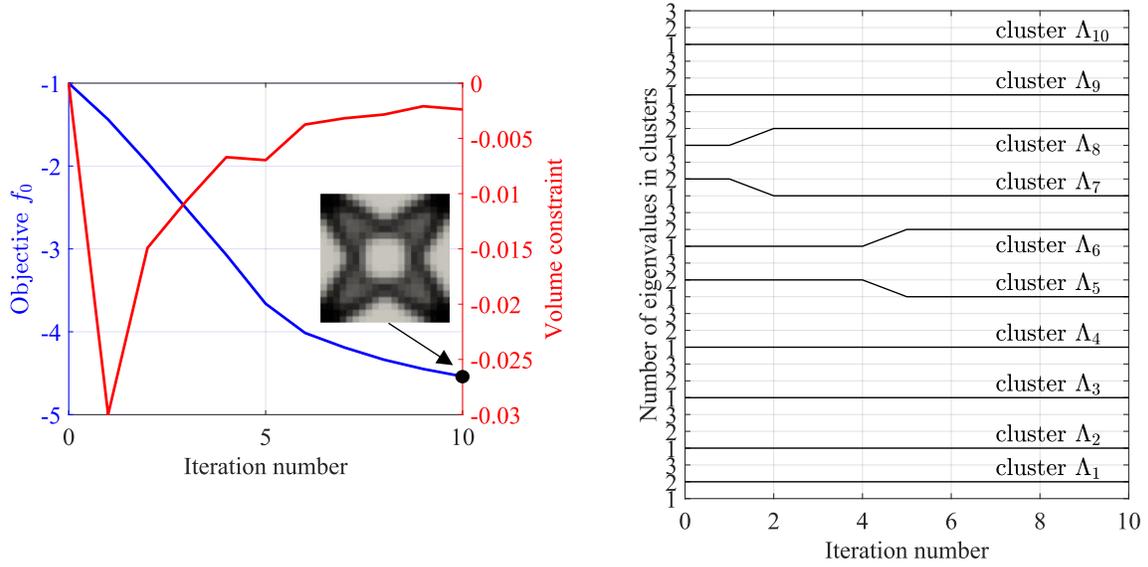

Figure 2. Square block fundamental eigen-cluster mean $\overline{\Lambda}_1$ maximization convergence history (left) and the eigen-cluster history (right) $\Lambda_k$, $k \in \{1, 6, 8\}$ has 2 repeated eigenvalues each (½-symmetry) at iteration 10.

Table 1. Eigen-clusters and associated eigenvalues (½-symmetry) at iteration 10.

---

$\Lambda_1$: {$\lambda_1$= 173584.09242419235, $\lambda_2$= 173584.0924259618}; $\Lambda_2$: {$\lambda_3$= 175053.68053465648}

$\Lambda_3$: {$\lambda_4$= 319883.38315195515}; $\Lambda_4$: {$\lambda_5$= 546645.0391030415}

$\Lambda_5$: {$\lambda_6$= 573905.3943252093}; $\Lambda_6$: {$\lambda_7$= 624466.5561570638, $\lambda_8$= 624466.5561571803}

$\Lambda_7$: {$\lambda_9$= 897071.7599316789}; $\Lambda_8$: {$\lambda_{10}$= 1172273.3071195404, $\lambda_{11}$= 1172273.307120035}

$\Lambda_9$: {$\lambda_{12}$= 1374885.2674479703}; $\Lambda_{10}$: {$\lambda_{13}$= 1733852.835891627}

---

The sensitivity results for the individual multiple eigenvalues under the ½-symmetry condition are presented in Figure 3 and Figure 4. The mismatch between the analytical sensitivities and the sensitivities calculated by CDM in Figure 3 demonstrates that multiple eigenvalues are not differentiable w.r.t all design variables. Moreover, the sensitivity result in Figure 4 shows that the multiple eigenvalues are also not differentiable w.r.t symmetric design variables. Thus, for this example, the multiple eigenvalues are not differentiable, and neither the analytical sensitivities nor the sensitivities by CDM are correct. Symmetric functions of multiple eigenvalues can be employed to resolve the non-differentiability of repeated eigenvalues, as demonstrated next.

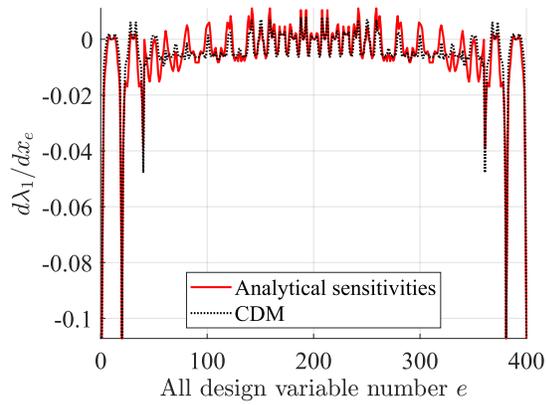
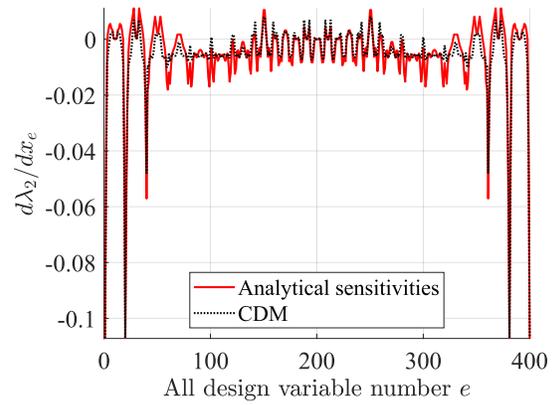

(a) $\frac{d\lambda_1}{dx_e}$          (b) $\frac{d\lambda_2}{dx_e}$

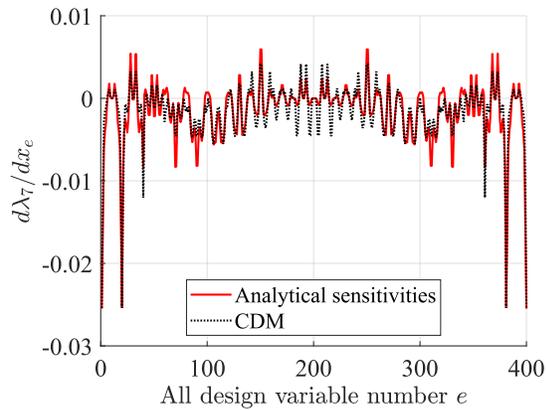
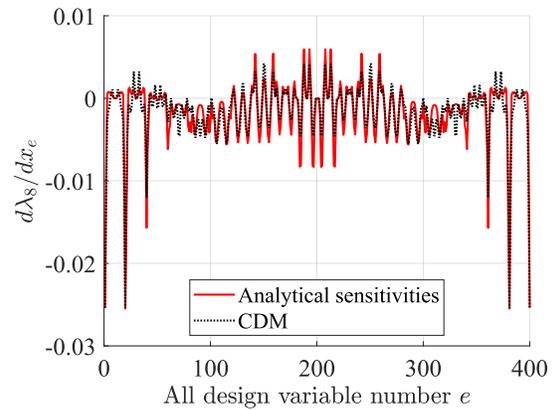

(c) $\frac{d\lambda_7}{dx_e}$          (d) $\frac{d\lambda_8}{dx_e}$

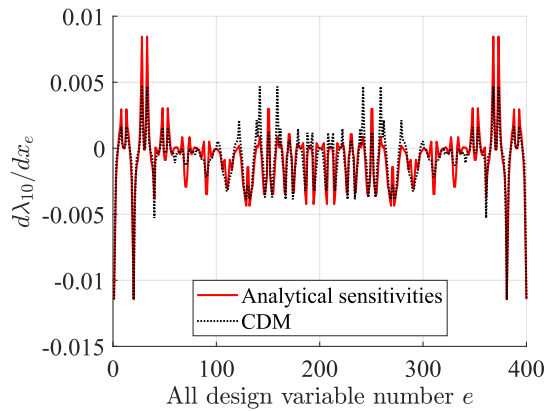
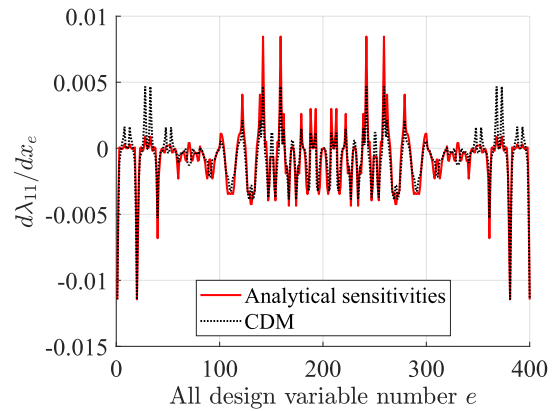

(e) $\frac{d\lambda_{10}}{dx_e}$          (f) $\frac{d\lambda_{11}}{dx_e}$

Figure 3. Sensitivities of multiple eigenvalues $\Lambda_k$, $k \in \{1, 2, 7, 8, 10, 11\}$ w.r.t *all* design variables at iteration 10 (½-symmetry).

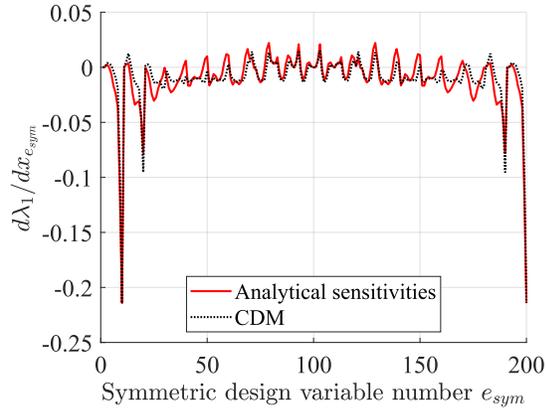
(a) $\frac{d\lambda_1}{dx_{e_{sym}}}$

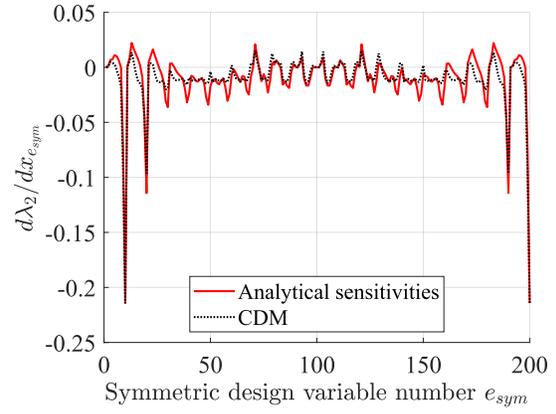
(b) $\frac{d\lambda_2}{dx_{e_{sym}}}$

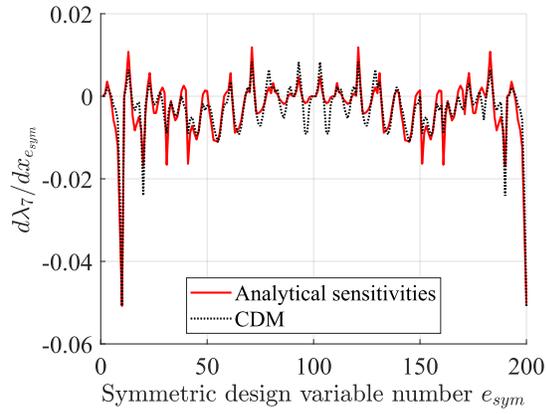
(c) $\frac{d\lambda_7}{dx_{e_{sym}}}$

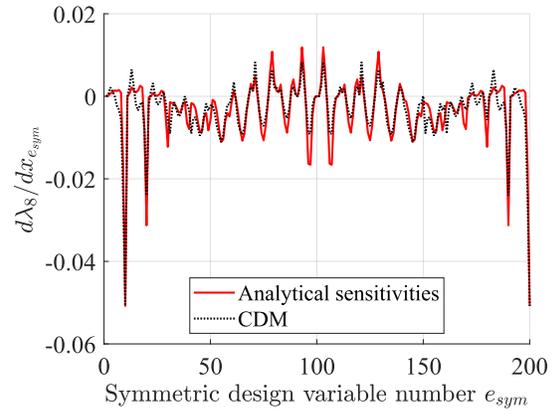
(d) $\frac{d\lambda_8}{dx_{e_{sym}}}$

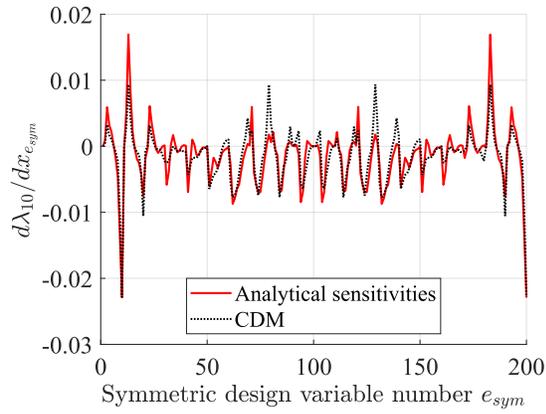
(e) $\frac{d\lambda_{10}}{dx_{e_{sym}}}$

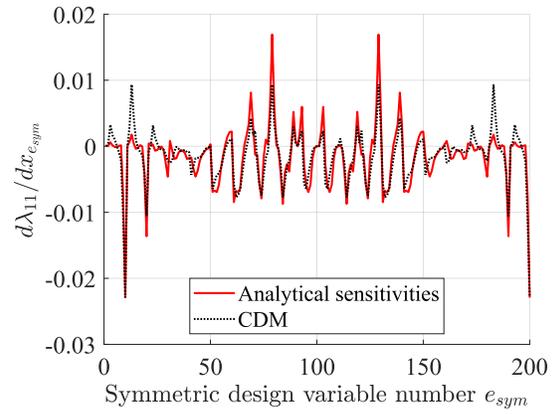
(f) $\frac{d\lambda_{11}}{dx_{e_{sym}}}$

Figure 4. Sensitivities of multiple eigenvalues $\Lambda_k$, $k \in \{1, 2, 7, 8, 10, 11\}$ w.r.t *symmetric* design variables at iteration 10 (½-symmetry).

*2.4.1.1 Cluster mean function*

Sensitivity analysis results of the cluster mean function (Eq. (14)) using Eq. (20) and CDM w.r.t all design variables and w.r.t symmetric design variables are shown in Figure 5 and Figure 6, respectively. These results show a good match between the analytical sensitivities and the sensitivities calculated by CDM. This verifies that the mean functions ($\bar{\Lambda}_1, \bar{\Lambda}_6$ and $\bar{\Lambda}_8$) of the corresponding eigen-clusters $\Lambda_1$, $\Lambda_6$ and $\Lambda_8$ which contain the multiple eigenvalues $\lambda_k$, $k \in \{1, 2, 7, 8, 10, 11\}$ are differentiable w.r.t all design variables and symmetric design variables when ½-symmetry is enforced. These multiple eigenvalues without clustering, however, are not differentiable w.r.t either all design variables (Figure 3) or symmetric design variables (Figure 4). The following two test cases illustrate the differentiability of $p$-norm and KS functions when appropriate clusters are included in these symmetric functions.

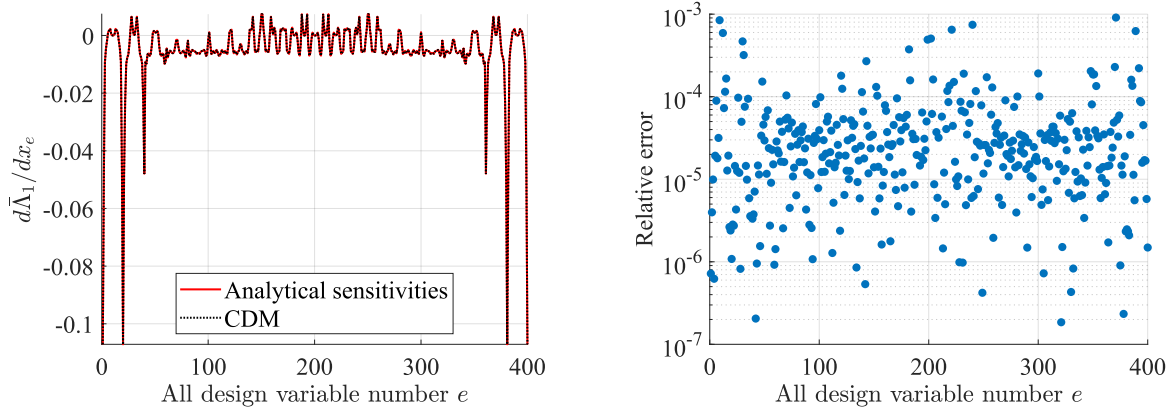

(a) $\dfrac{d\bar{\Lambda}_1}{dx_e}$ and relative error

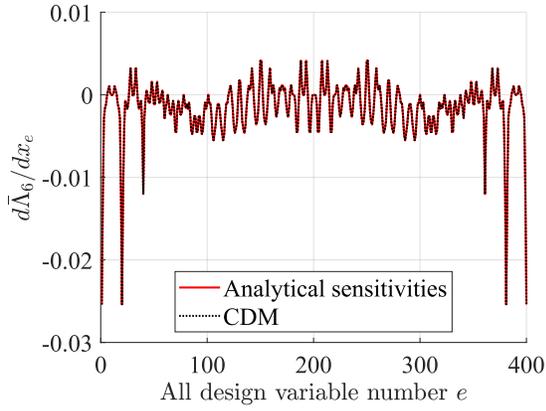
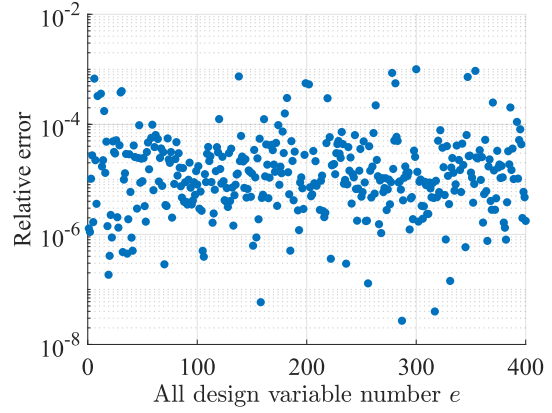

(b) $\frac{d\bar{\Lambda}_6}{dx_e}$ and relative error

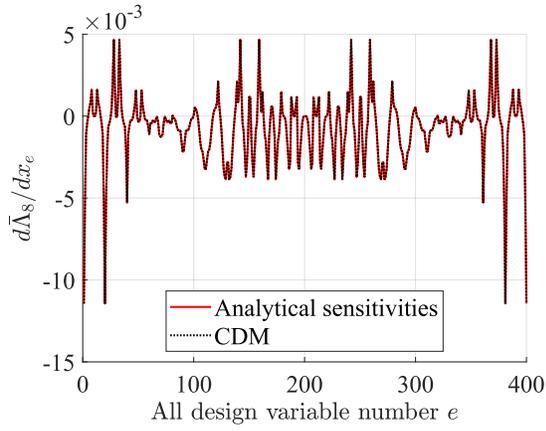
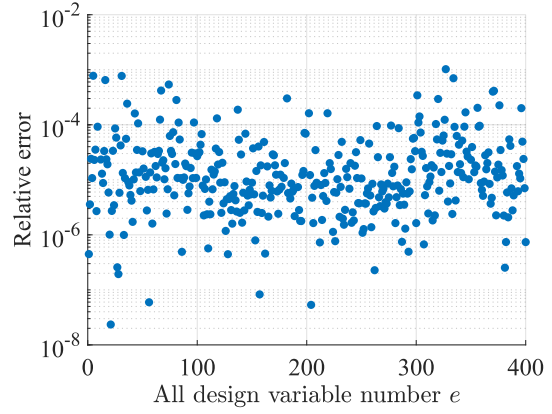

(c) $\frac{d\bar{\Lambda}_8}{dx_e}$ and relative error

Figure 5. Verification of sensitivities of the cluster mean of the eigen-clusters with multiple eigenvalues $\Lambda_k$, $k \in \{1,6,8\}$ w.r.t *all* design variables (½-symmetry) at iteration 10.

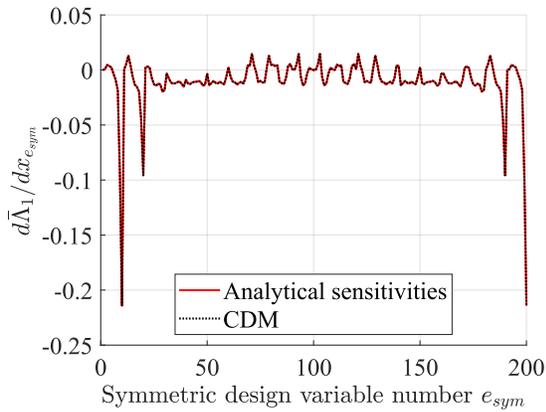
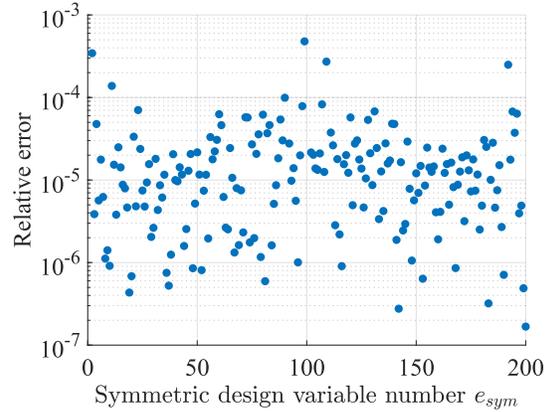

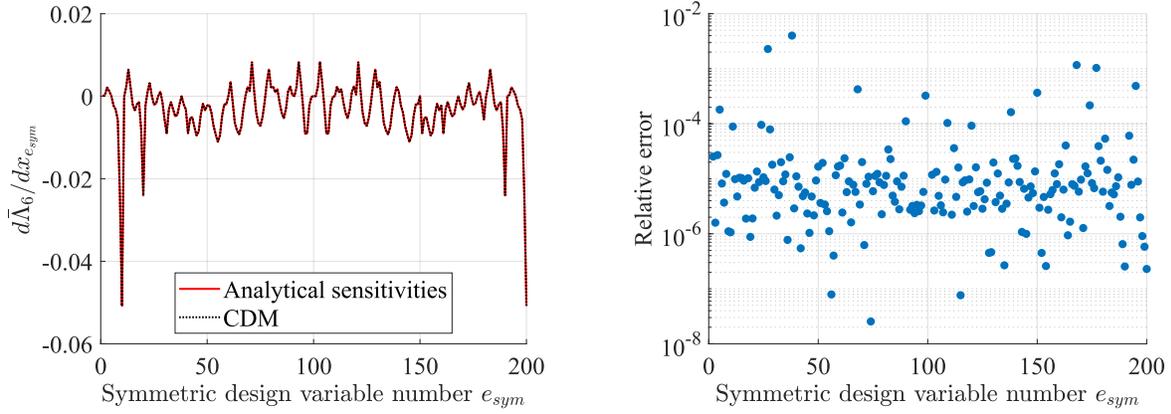

(b) $\dfrac{d\bar{\Lambda}_6}{dx_{e_{sym}}}$ and relative error

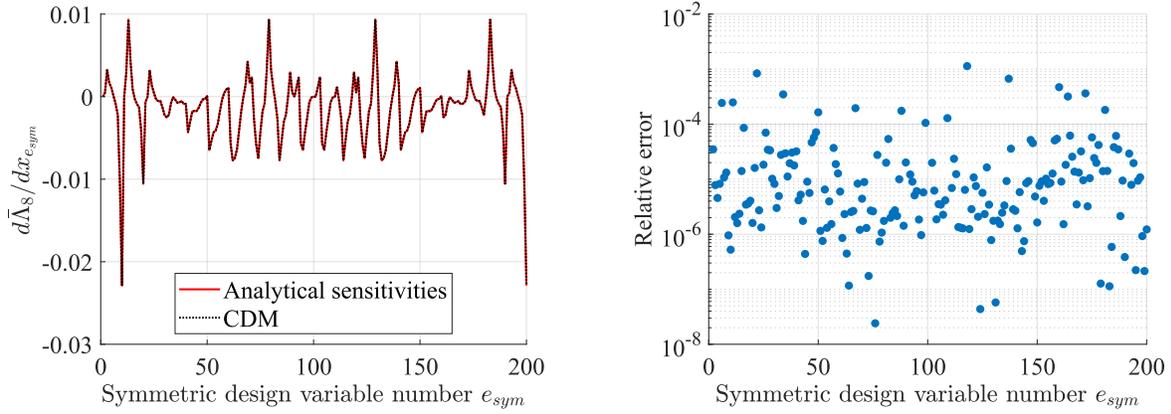

(c) $\dfrac{d\bar{\Lambda}_8}{dx_{e_{sym}}}$ and relative error

Figure 6. Verification of sensitivities of the cluster mean of the eigen-clusters with multiple eigenvalues $\Lambda_k$, $k \in \{1,6,8\}$ w.r.t *symmetric* design variables (½-symmetry) at iteration 10.

*2.4.1.2 p-norm function*

As a smooth approximation of the maximum norm, $\|\blacksquare\|_\infty$, the $p$-norm function $\|\blacksquare\|_p$ is frequently used in eigenvalue topology optimization. For instance, the smooth approximation of the largest eigenvalue $\lambda^*$ of the first $n$ number of eigenvalues constructed through the utilization of the $p$-norm function is outlined below.

$$\lambda^* = \|\boldsymbol{\lambda}\|_p = \left(\sum_{k=1}^{n} \lambda_k^p\right)^{1/p} \tag{21}$$

where $\lambda_{max} = \lim_{p \to \infty} \|\boldsymbol{\lambda}\|_p = \|\boldsymbol{\lambda}\|_\infty$. A numerically stable formulation of the $p$-norm function approximating the maximum of the first $n$ eigenvalues and the associated derivative are expressed as

$$\lambda^* = \|\boldsymbol{\lambda}\|_p = \beta_0 \left[\sum_{k=1}^{n} \left(\frac{\lambda_k}{\beta_0}\right)^p\right]^{1/p}$$

$$\frac{d\lambda^*}{d\lambda_k} = \left[\sum_{k=1}^{n} \left(\frac{\lambda_k}{\beta_0}\right)^p\right]^{\frac{1}{p}-1} \left(\frac{\lambda_k}{\beta_0}\right)^{p-1} \tag{22}$$

where $\beta_0$ is a constant, e.g., set $\beta_0 = \max(\boldsymbol{\lambda})$.

In the context of the square block example, consider the $p$-norm of the first 8 eigen-clusters. A total of 11 eigenvalues are included in the first 8 eigen-clusters, and $\Lambda_8$ has repeated eigenvalues ($\Lambda_8 = \{\lambda_{10}, \lambda_{11}\}$). The $p$-norm function, when constructed with the inclusion of *all* the eigenvalues in the specified first $n$ eigen-clusters, is a symmetric polynomial of multiple eigenvalues and thus differentiable. Therefore, to construct a differentiable approximation of the maximum eigenvalue among the first 8 eigen-clusters, a total of 11 eigenvalues ($\boldsymbol{\lambda} = [\lambda_1, \dots, \lambda_{11}]$) must be included in the computation of $\|\boldsymbol{\lambda}\|_p$. The $p$-norm function, without consideration of clustering, does not guarantee differentiability. As an example, when $\lambda_{11}$ is not included, that is, when $\Lambda_8$ is not a complete cluster anymore, the $p$-norm function of the first 10 eigenvalues is no longer a symmetric function of multiple eigenvalues and hence is *not* differentiable. The mismatch between the analytical sensitivities and the CDM results is illustrated in Figure 7 in the case of the incomplete eigen-cluster. Figure 8 shows the sensitivities of the $p$-norm function, encompassing all 11 eigenvalues across the first 8 eigen-clusters. When the eigenvalues $\lambda_{10}$ and $\lambda_{11}$ in the last eigen-cluster $\Lambda_8$ are both included in $\boldsymbol{\lambda}$ to compute $\|\boldsymbol{\lambda}\|_p$, the analytical sensitivities match the CDM results. This verifies that when all the eigenvalues in eigen-clusters are included, the corresponding $p$-norm of the considered eigenvalues becomes differentiable. Therefore, to ensure differentiability, when the $p$-norm function is utilized to approximate the extremum of the computed eigenvalues, the relevant eigen-clusters must be complete.

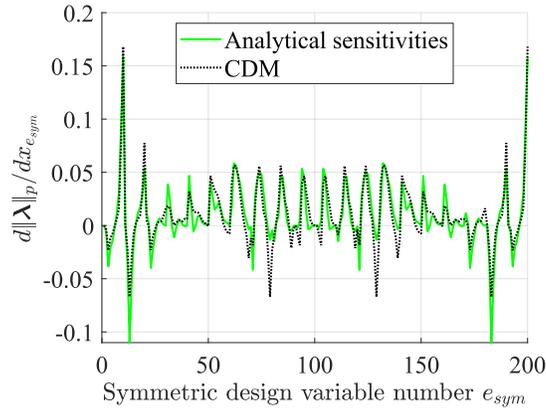

Figure 7. $p$-norm sensitivities w.r.t *symmetric* design variables ($p$ =10) (½-symmetry enforced) at iteration 10 when the last eigen-cluster is *incomplete*.

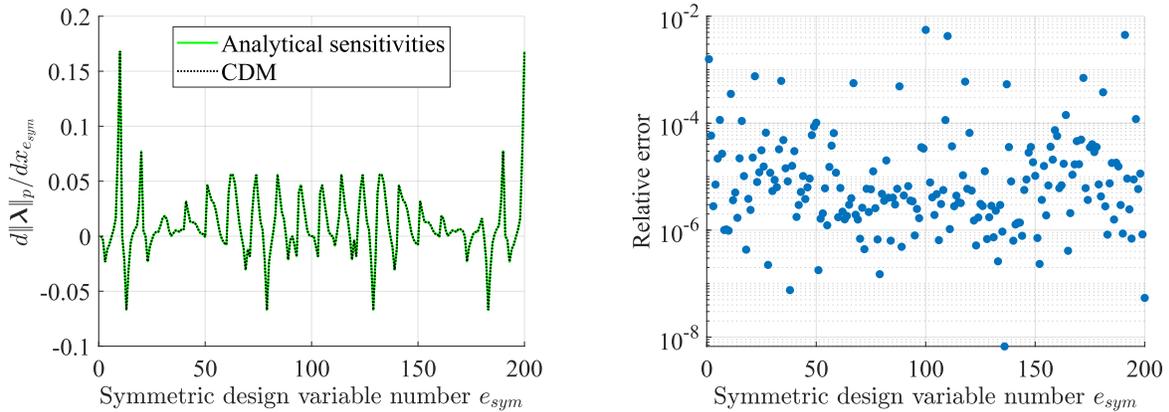

Figure 8. $p$-norm sensitivities w.r.t *symmetric* design variables ($p$ =10) (½-symmetry) at iteration 10 when the first 8 eigen-clusters are *complete*.

### 2.4.1.3 Kreisselmeier-Steinhauser (KS) function

The Kreisselmeier-Steinhauser (KS) function is another smooth function that is commonly used in optimization to approximate the extremum of a set of values. A smooth approximation of the largest eigenvalue $\lambda^*$ of the first $n$ eigenvalues using the KS function is expressed as

$$\lambda^* = KS(\pmb{\lambda}, q) = \frac{1}{q}\ln\left(\sum_{k=1}^{n}\exp(q\lambda_k)\right) \quad (23)$$

and $\lambda_{max} = \lim_{q\to\infty}\lambda^*$. A numerically stable formulation of the KS function approximating the maximum of $n$ number of eigenvalues and the associated derivative are given by

$$\lambda^* = KS(\pmb{\lambda}, q) = \beta_0 + \frac{1}{q}\ln\left(\sum_{k=1}^{n}\exp[q(\lambda_k - \beta_0)]\right)$$

$$\frac{d\lambda^*}{d\lambda_k} = \frac{\exp[q(\lambda_k - \beta_0)]}{\sum_{k=1}^{n}\exp[q(\lambda_k - \beta_0)]} \quad (24)$$

where $\beta_0$ is a constant, e.g., set $\beta_0 = \max(\pmb{\lambda})$.

Like the case of $p$-norm function, the extremum approximation of an arbitrary set of eigenvalues by the KS function does not ensure differentiability. To guarantee differentiability, eigenvalues must be clustered first. Once the eigen-clusters are established, the KS approximation of *all* eigenvalues contained in the *complete* set of eigen-clusters becomes differentiable. Figure 9 and Figure 10 illustrate the KS function sensitivity results with incomplete and complete eigen-cluster $\Lambda_8$, respectively. The close match between the analytical sensitivities and those calculated by CDM (Figure 10) confirms that the KS function with complete eigen-clusters is a differentiable function of multiple eigenvalues.

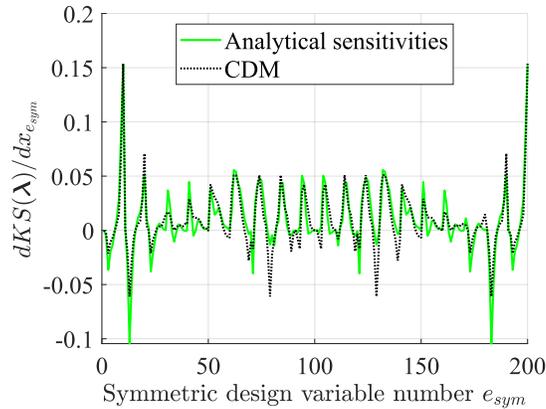

Figure 9. KS function sensitivities w.r.t *symmetric* design variables ($q=10$) (½-symmetry enforced) at iteration 10 when the last eigen-cluster is *incomplete*.

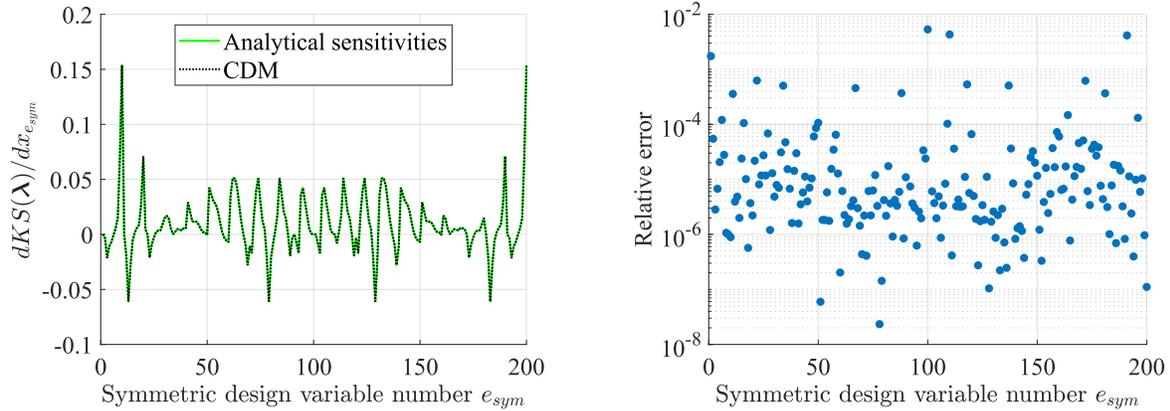

Figure 10. KS sensitivities w.r.t *symmetric* design variables ($q = 10$) (½-symmetry enforced) at iteration 10 and relative error when the first eight eigen-clusters are *complete*.

### 2.4.2 Symmetry Type-2: ⅛ -symmetry

An additional study with increased symmetry is carried out to further illustrate the connection between the underlying symmetry and the sensitivities of multiple eigenvalues. To this end, ⅛-symmetry is enforced in the square block design domain (Figure 11) and sensitivity analysis is carried out on a design obtained after 10 iterations, where again a total of 10 eigen-clusters are included in the constraints in optimization formulation Eq. (15). The intermediate design after 10 iterations and convergence history are shown in Figure 12. The eigen-cluster history shown in Figure 12 reveals that eigen-clusters $\Lambda_1$, $\Lambda_6$ and $\Lambda_8$ each contains two repeated eigenvalues. The detailed information of the eigen-clusters at iteration step 10 is summarized in Table 2.

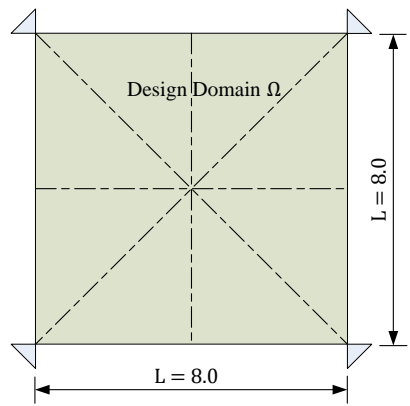

Figure 11. Square block design domain with dashed lines indicating the axes of symmetry (⅛-symmetry).

Page 29 of 73

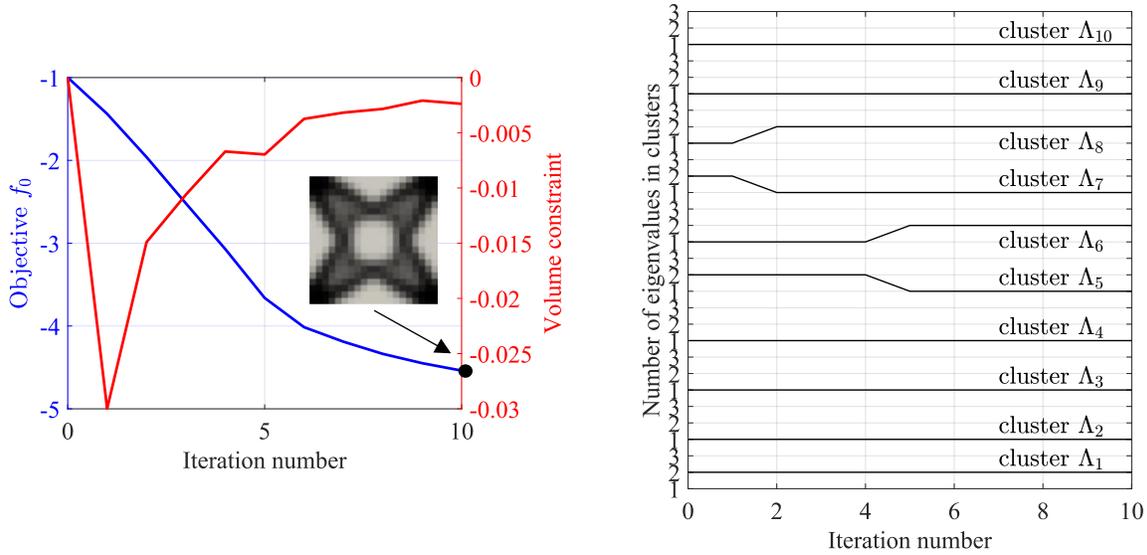

Figure 12. Square block fundamental eigen-cluster mean $\overline{\Lambda}_1$ maximization convergence history (left) and the eigen-cluster history (right) $\Lambda_k$, $k \in \{1, 6, 8\}$ has 2 repeated eigenvalues each (⅛-symmetry) at iteration 10.

Table 2. Eigen-clusters and associated eigenvalues (⅛-symmetry) at iteration 10.

| |
|---|
| $\Lambda_1$: {$\lambda_1$=173653.26400813632, $\lambda_2$=173653.26400813943}; $\Lambda_2$: {$\lambda_3$=175131.2530074806} |
| $\Lambda_3$: {$\lambda_4$=320061.37122083077}; $\Lambda_4$: {$\lambda_5$=546678.5142011326} |
| $\Lambda_5$: {$\lambda_6$=573940.5095230478}; $\Lambda_6$: {$\lambda_7$=624572.5471291995, $\lambda_8$=624572.5471292019} |
| $\Lambda_7$: {$\lambda_9$=897542.8745201216}; $\Lambda_8$: {$\lambda_{10}$=1172840.394289676, $\lambda_{11}$=1172840.3942896815} |
| $\Lambda_9$: {$\lambda_{12}$=1375569.92504824}; $\Lambda_{10}$: {$\lambda_{13}$=1734117.185796525} |

Sensitivity analysis is carried out using the analytical method and CDM w.r.t all design variables and w.r.t symmetric design variables. The sensitivity results are presented in Figure 13 and Figure 14, respectively. Based on the results shown in Figure 13, it is clear that the multiple eigenvalues $\lambda_k$, $k = \{1, 2, 7, 8, 10, 11\}$ are not differentiable w.r.t *all* design variables when ⅛-symmetry is enforced. However, Figure 14 shows that the multiple eigenvalues *are differentiable* w.r.t the *symmetric* design variables corresponding to the enforced ⅛-symmetry. Thus, in this higher symmetry case, even the tested multiple eigenvalues are differentiable. This result with ⅛-symmetry is in contrast with the ½-symmetry case where the multiple eigenvalues are non-differentiable w.r.t symmetric design variables (Figure 4). This result also confirms the study in Ref. [48], which shows that under certain symmetry conditions, repeated eigenvalues can be differentiable.

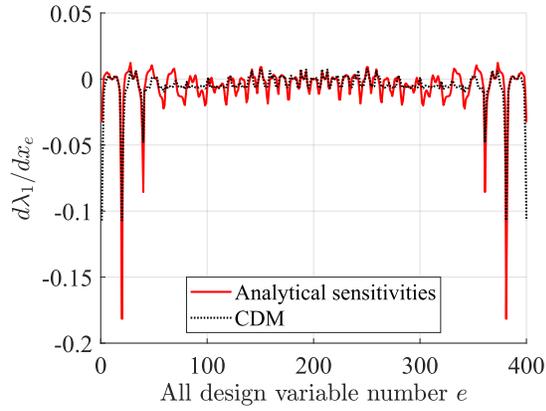

(a) $\frac{d\lambda_1}{dx_e}$

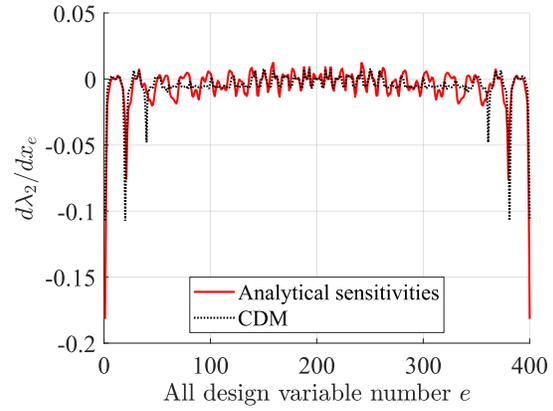

(b) $\frac{d\lambda_2}{dx_e}$

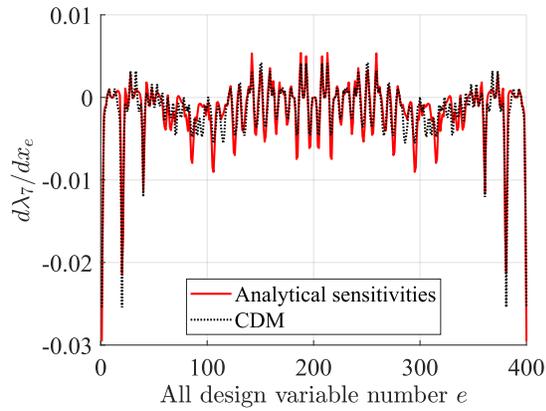

(c) $\frac{d\lambda_7}{dx_e}$

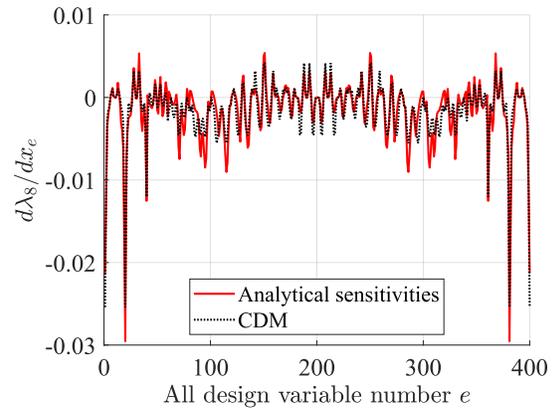

(d) $\frac{d\lambda_8}{dx_e}$

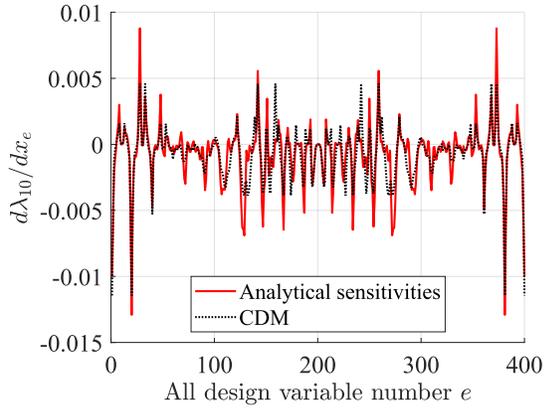

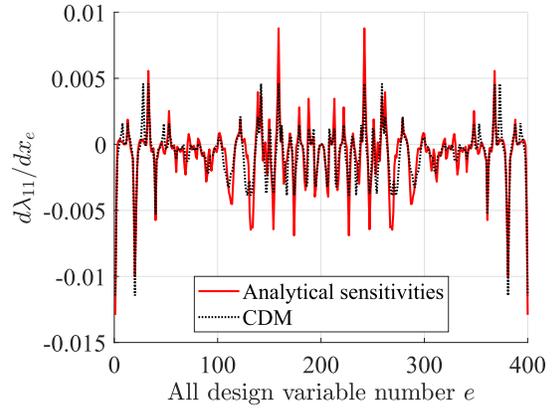

(e) $\frac{d\lambda_{10}}{dx_e}$  (f) $\frac{d\lambda_{11}}{dx_e}$

Figure 13. Sensitivities of multiple eigenvalues $\lambda_k$, $k \in \{1, 2, 7, 8, 10, 11\}$ w.r.t *all* design variables (⅛-symmetry).

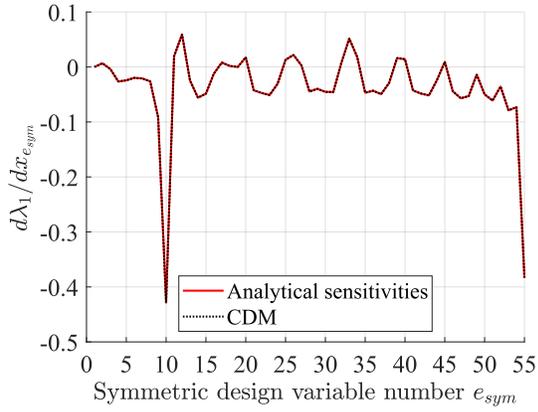

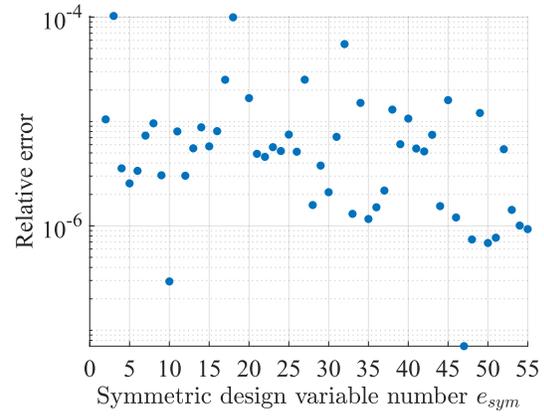

(a) $\frac{d\lambda_1}{dx_{e_{sym}}}$ and relative error

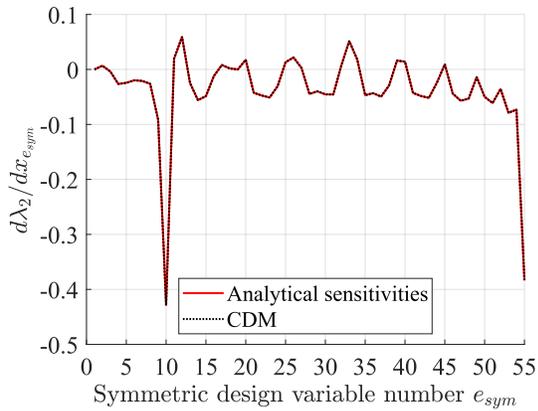

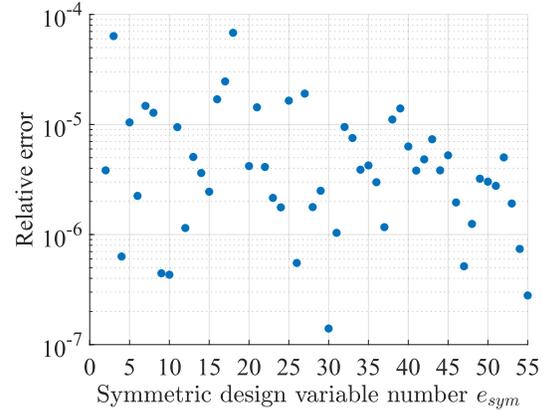

(b) $\frac{d\lambda_2}{dx_{e_{sym}}}$ and relative error

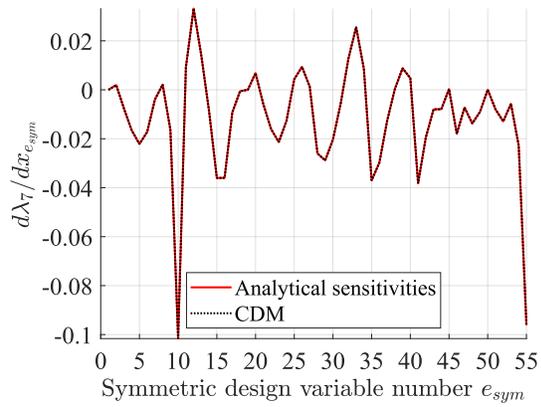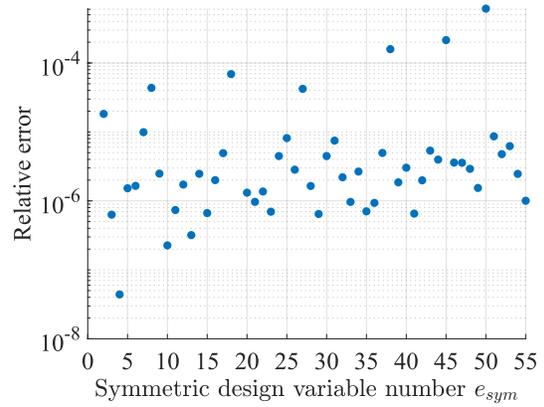

(c) $\frac{d\lambda_7}{dx_{e_{sym}}}$ and relative error

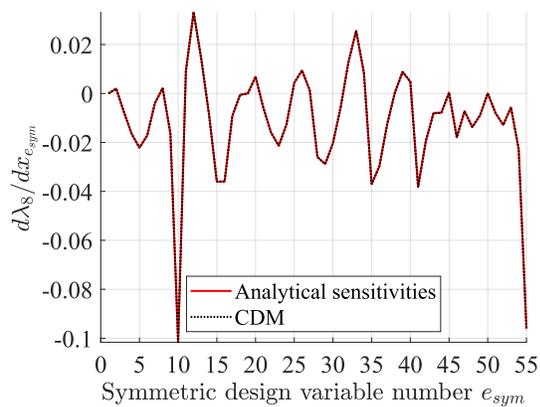
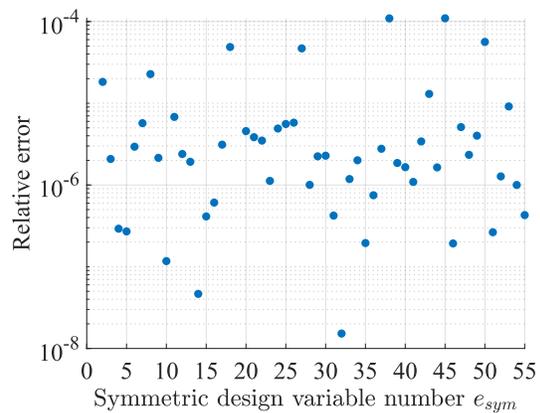

(d) $\frac{d\lambda_8}{dx_{e_{sym}}}$ and relative error

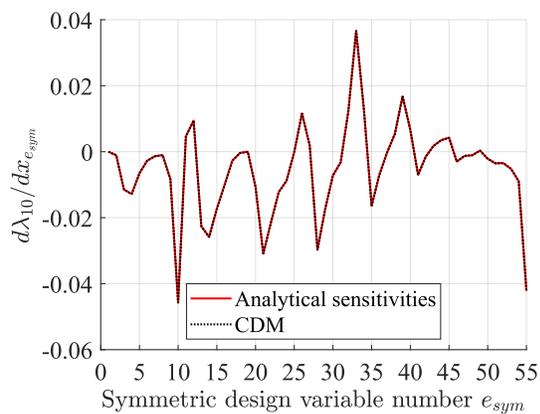
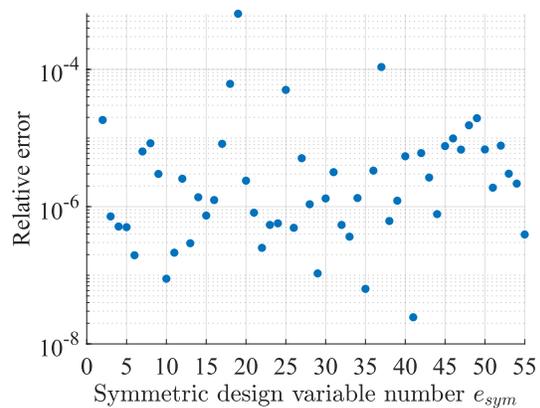

(e) $\frac{d\lambda_{10}}{dx_{e_{sym}}}$ and relative error

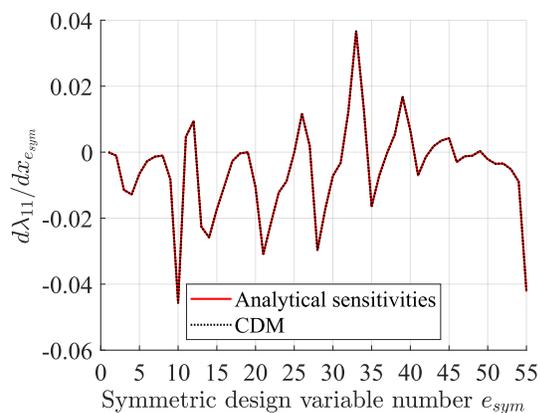
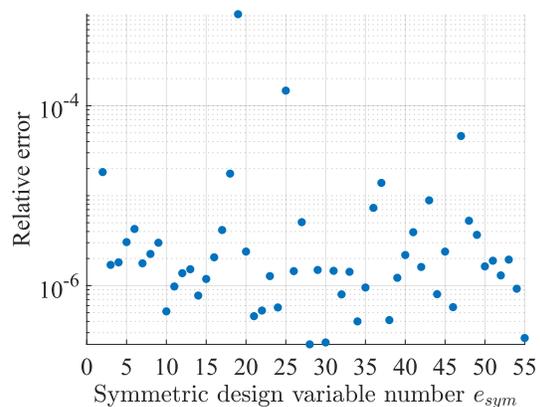

(f) $\frac{d\lambda_{11}}{dx_{e_{sym}}}$ and relative error

Figure 14. Verification of sensitivities of multiple eigenvalues $\lambda_k$, $k \in \{1,2,7,8,10,11\}$ w.r.t *symmetric* design variables (⅛-symmetry).

*2.4.2.1 Cluster mean function*

Next, the analytical sensitivities of cluster means $\overline{\Lambda}_1$, $\overline{\Lambda}_6$ and $\overline{\Lambda}_8$ are juxtaposed with those obtained through CDM for comparison in Figure 15 and Figure 16. When multiple eigenvalues are clustered with the mean function, the analytical sensitivities match those calculated by CDM. These results demonstrate that $\overline{\Lambda}_1$, $\overline{\Lambda}_6$ and $\overline{\Lambda}_8$ are differentiable w.r.t *all* design variables (Figure 15) and *symmetric* design variables (Figure 16) as well when the ⅛-symmetry is enforced.

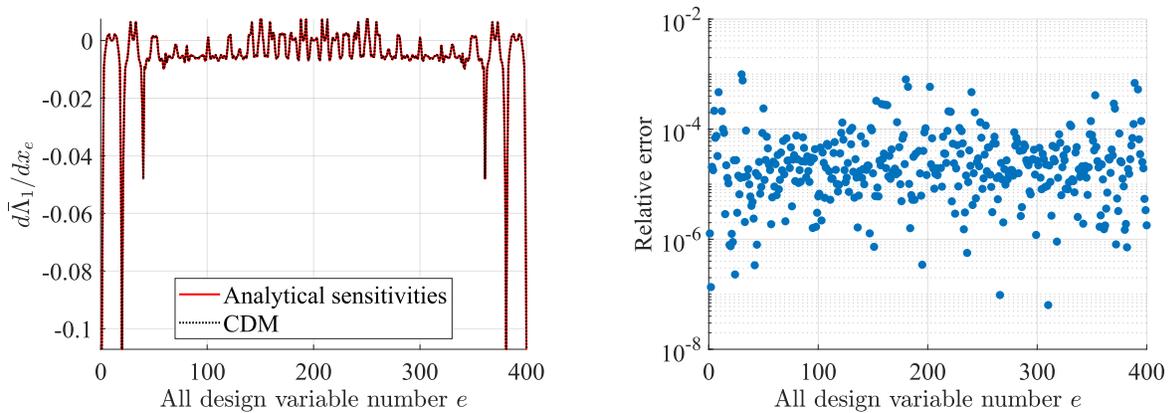

(a) $\frac{d\overline{\Lambda}_1}{dx_e}$ and relative error

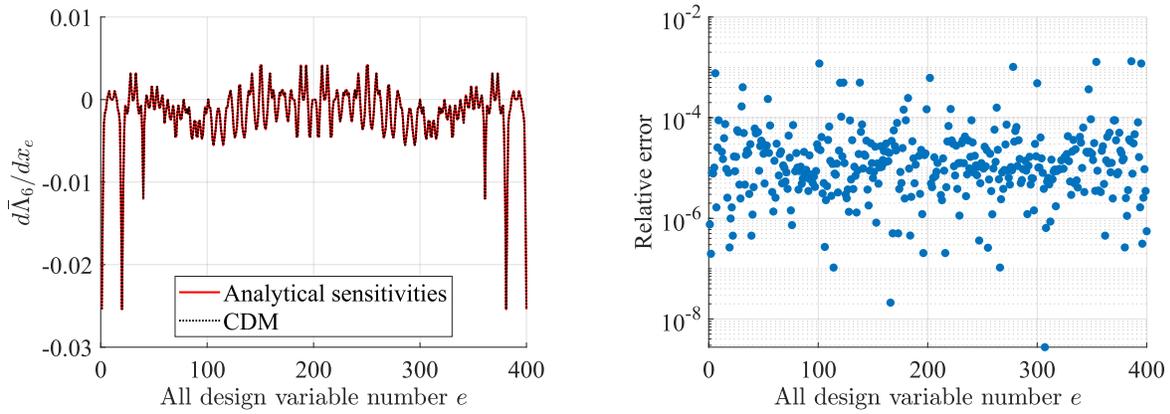

(b) $\frac{d\overline{\Lambda}_6}{dx_e}$ and relative error

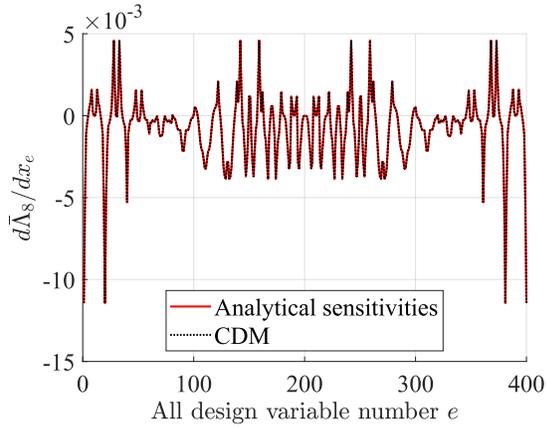
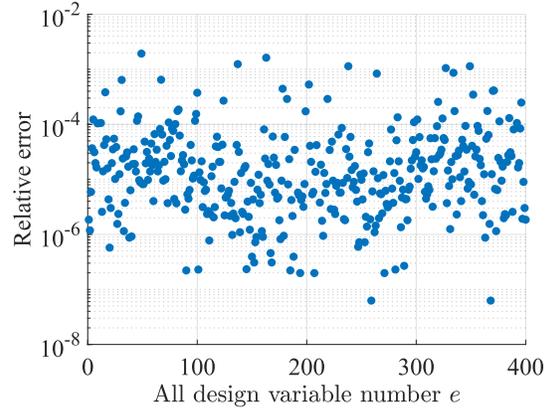

(c) $\frac{d\bar{\Lambda}_8}{dx_e}$ and relative error

Figure 15. Verification of sensitivities of the cluster mean of the eigen-clusters with multiple eigenvalues $\bar{\Lambda}_k$, $k \in \{1, 6, 8\}$ w.r.t all design variables (⅛-symmetry).

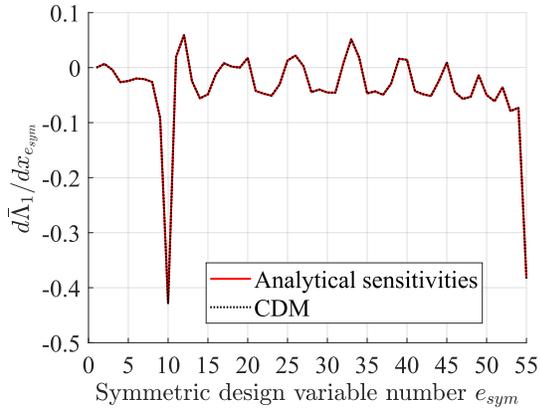
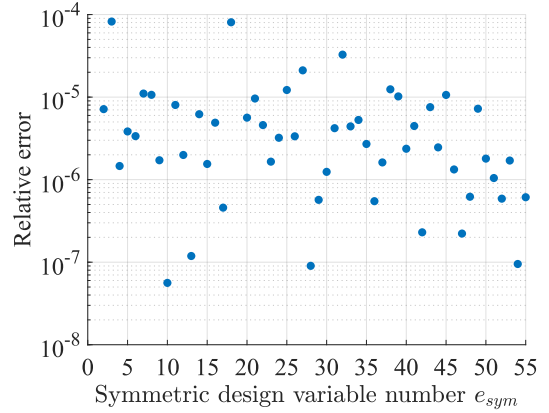

(a) $\frac{d\bar{\Lambda}_1}{dx_{e_{sym}}}$ and relative error

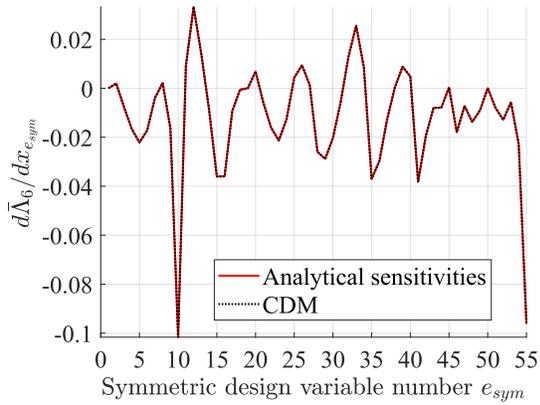
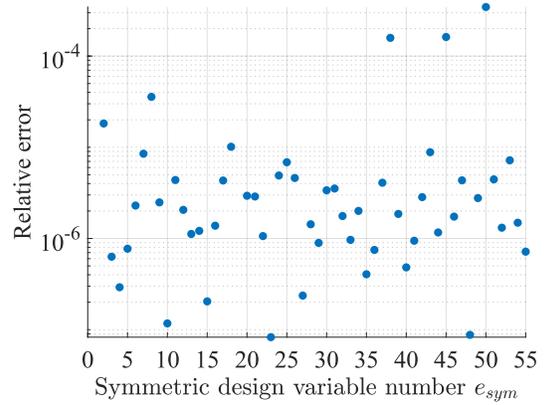

(b) $\frac{d\bar{\Lambda}_6}{dx_{e_{sym}}}$ and relative error

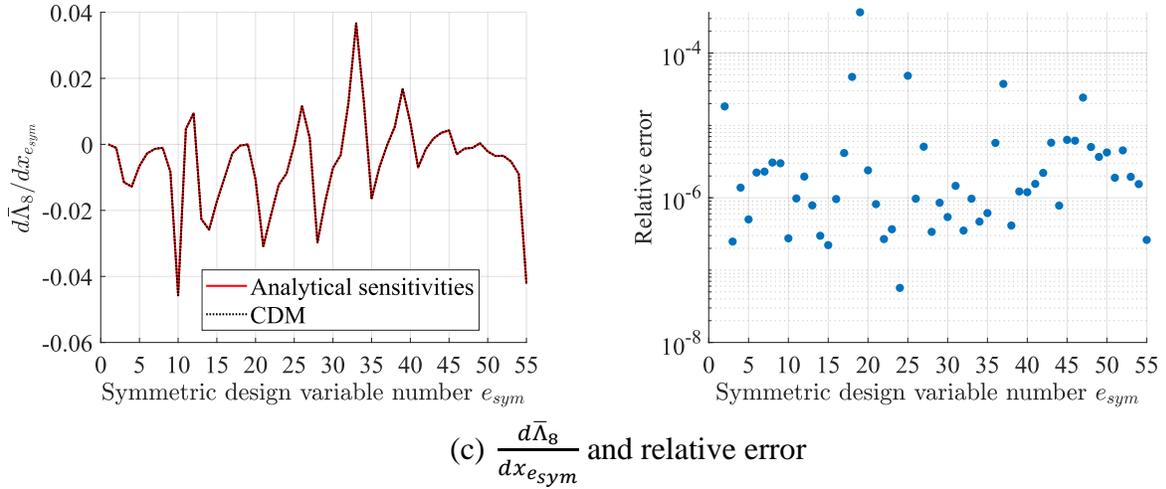

(c) $\dfrac{d\bar{\Lambda}_8}{dx_{e_{sym}}}$ and relative error

Figure 16. Verification of sensitivities of the cluster mean of the eigen-clusters with multiple eigenvalues $\bar{\Lambda}_k$, $k \in \{1,6,8\}$ w.r.t symmetric design variables (⅛-symmetry).

*2.4.3 Remarks*

With two enforced symmetries – i.e., ½ and ⅛ symmetry – the eigen-cluster means of multiple eigenvalues are shown to be differentiable w.r.t the *symmetric* design variables for both symmetries (Figure 6 and Figure 16). In addition, the cluster means are also differentiable w.r.t *all* design variables in both cases (Figure 5 and Figure 15). The individual multiple eigenvalues, however, are only differentiable w.r.t the *symmetric* design variables under ⅛-symmetry (Figure 14) but *not* differentiable under ½-symmetry (Figure 4). Moreover, the multiple eigenvalues without clustering are *not* differentiable w.r.t *all* design variables (Figure 3 and Figure 13). This exercise confirms that the cluster mean of multiple eigenvalues varies smoothly w.r.t the design variables. Thus, the mean function with clustering is always a differentiable function of multiple eigenvalues. Therefore, it is expedient to use cluster means – regardless of the enforced symmetry – to ensure the differentiability of multiple eigenvalues. Similarly, $p$-norm and KS functions are differentiable only if all the repeated eigenvalues are included. Also, note that in [49], the authors constructed a "mean-average" function of the eigenvalues without adequate consideration of eigenvalue multiplicities. Without clustering, the proposed mean-average is not a symmetric function of repeated eigenvalues, and thus, differentiability is not guaranteed.

## 2.5 Optimization Procedure

The overall optimization workflow using the cluster mean approach is presented in Table 3 below.

Table 3. Workflow of eigenfrequency topology optimization employing the cluster mean approach.

| Optimization Algorithm using Cluster Mean Approach |
|---|
| a) Initialize design variables. |
| b) Apply density filters as applicable (Eq. (7)) and interpolate the density variables to obtain mass and stiffness matrices. |
| c) Solve for the first $N$ smallest eigenvalues of the system $(\lambda_1, \dots, \lambda_N)$ (Eq. (1)). |
| d) Eigenvalue clustering operation:<br>  1) Create the required $m$ clusters.<br>  2) Compute extra eigenvalues if $m$ number of clusters requires more than the $N$ eigenvalues.<br>  3) Calculating the mean value of each cluster (Eq. (14)). |
| e) Calculate the objective function, constraints, and their sensitivities. |
| f) Use the MMA optimizer to update design variables. |
| g) Return to (b) until the desired number of optimization iterations or another specified termination criterion is reached. |

## 3 Numerical Results

In this section, the efficacy of the proposed bound formulations using cluster means in various optimization problems is demonstrated through a series of numerical results that include maximization of the fundamental eigen-cluster mean $\bar{\Lambda}_1$, the $n^{th}$-order eigen-cluster mean $\bar{\Lambda}_n$ and the $m^{th}$-order bandgap $(\bar{\Lambda}_{m+1} - \bar{\Lambda}_m)$. For solid-void cases, the initial design is homogeneous with the design variables initialized to volume fraction threshold, i.e., $x_e^{(0)} = V^f$. The initial design of the multi-material with void phase cases is also homogeneous. In this case, the $k^{th}$ solid phase design variable $x_{k,e}^{(0)}$ is initialized to $x_{k,e}^{(0)} = \begin{cases} V_1^f & k=1 \\ V_2^f/V_1^f & k=2 \end{cases}$ and $x_{k,e}^{(0)} = \begin{cases} V_1^f & k=1 \\ V_2^f/V_1^f & k=2 \\ V_3^f/V_2^f & k=3 \end{cases}$ for bi-material + void phase and tri-material + void phase, respectively. The threshold $\rho_T$ in Eqns. (8), (11), and (13) are set to 0.1 for 2D and plate problems and 0.02 for 3D problems. A total number of 500 optimization iterations are carried out for all the test cases. For computational implementation, an in-house object-oriented Python-based library is utilized for finite element eigen-analysis and optimization tasks. The most computationally intensive task during the optimization process is the eigenvalue analysis. To accomplish this, a SciPy eigen-solver [50] that uses the ARPACK library [51] is employed, which can effectively calculate the desired spectrum. Moreover, to compute the first $N$-smallest eigenvalues, the shift-inverse operation is first performed as ARPACK is effective for extracting the largest spectrum. As the shift-inverse operation involves solving a linear system, the PARDISO solver [52] is employed to solve the underlying sparse linear system in the shift-inverse operations. The Center for Research Computing (CRC) at the University of Notre Dame provided the computing resources utilized for result generation. The computer hardware includes a dual 12-core Intel(R) Xeon(R) CPU E5-2680 v3 @ 2.50GHz Haswell processor and 256 GB RAM.

### 3.1  2D Examples

#### 3.1.1  Square Block

The square block design domain in Figure 11 is discretized by 400×400 4-node quadrilateral elements. Poisson's ratio is 0.3, the density filter radius is 0.06, and the volume fraction is restricted to 0.48. Notably, the design space admits ⅛-symmetry and such symmetry is enforced in the design (Figure 11). The optimized designs of the square block for the first fundamental eigen-cluster mean $\bar{\Lambda}_1$, second eigen-cluster mean $\bar{\Lambda}_2$, and the bandgap ($\bar{\Lambda}_2 - \bar{\Lambda}_1$) and their corresponding convergence histories are shown in Figure 17. The objective value decreases smoothly as the optimization progresses. Figure 17(a) to (c) show that the volume constraints are satisfied but may not be active in the optimized design. This feature is similar to that reported in [29].

The eigenfrequencies and eigen-cluster histories are shown in Figure 18. The eigen-cluster plots record the number of repeated eigenfrequencies in each eigen-cluster throughout the optimization. It is important to note that the plots only show the first five clusters, whereas ten clusters are incorporated into the optimization constraints. Remarkably, relying solely on the *visual inspection* of eigenvalue history may lead to misleading conclusions when determining the multiplicity of specific eigenfrequencies. As an example, in Figure 18(a), eigenfrequency $\omega_3$ converges towards eigenfrequencies $\omega_1$ and $\omega_2$ around step 380, and later seem to merge completely. However, the eigen-cluster history reveals that the eigenfrequency $\omega_3$ remains simple, while eigenfrequencies $\omega_1$ and $\omega_2$ are repeated (i.e., belonging to the same eigen-cluster). Similar observations can be made in Figure 18(b) and (c), where the initially separated eigenfrequencies $\omega_3$, $\omega_4$ and $\omega_5$ become close but remain separate according to the eigen-cluster history. Thus, when relying solely on visual inspection of the eigenvalue history, it may be asserted that the optimization process transforms separate eigenvalues into a repeated eigenvalue with a multiplicity of 3, which is incorrect. Only eigenfrequencies $\omega_1$ and $\omega_2$ consistently maintain their repeated status throughout the entire optimization process. Based on the above observations, the eigenfrequencies are classified as (a) *simple* – if they remain separate; (b) *convergent* – if they come close but are still in different clusters; and (c) *repeated* – if they are in the same cluster with the clustering tolerance $\epsilon_{\text{tol}} = 1.0\text{E-}8$. Note that simple and convergent eigenvalues are differentiable, and no special treatment is needed for these eigenvalues.

Transitions in clusters are also observed in these problems, characterized by the fluctuations in the number of repeated eigenvalues between clusters, exhibiting brief increases or decreases. This can be attributed to the temporary numerical clustering of two convergent eigenvalues. Figure 19 illustrates a zoomed-in view of the eigen-cluster history shown in Figure 18(c)(right). In the enlarged view, a distinct change in the eigen-cluster is highlighted, showcasing the transition from optimization step 416 to step 417. For illustration, the corresponding eigenmodes at step 416 and step 417 are plotted in Figure 19. The eigen-clusters in step 416 are recorded as $\Lambda_3: \{\lambda_4 = \omega_4^2 = 751339.4804959776\}$, $\Lambda_4: \{\lambda_5 = \omega_5^2 = 751339.5183880392\}$, and $\Lambda_5: \{\lambda_6 = \omega_6^2 = 831409.3619366486, \lambda_7 = \omega_7^2 = 831409.3619371024\}$. Note that at step 416, $\lambda_4$ and $\lambda_5$ are two convergent eigenvalues and are grouped into separate eigen-clusters. On the other hand, $\lambda_6$ and $\lambda_7$ are classified as repeated. At the following step 417, the eigen-clusters are recorded as $\Lambda_3: \{\lambda_4 = \omega_4^2 = 751349.7695886609, \lambda_5 = \omega_5^2 = 751349.7738859111\}$, $\Lambda_4: \{\lambda_6 = \omega_6^2 = 831556.0332389948, \lambda_7 = \omega_7^2 = 831556.0332399898\}$, $\Lambda_5 = \{\lambda_8 = \omega_8^2 = 866334.587675779\}$. At step 417, $\lambda_6$ and $\lambda_7$ again remain in the same cluster. However, at this step, $\lambda_4$ and $\lambda_5$ also come close enough and are temporarily assigned to the same cluster. Interestingly, the optimizer separates $\lambda_4$ and $\lambda_5$ at the subsequent step, and they remain in separate clusters afterward. In contrast, $\lambda_6$ and $\lambda_7$ continue to stay clustered. With this insight, it is posited that $\lambda_6$ and $\lambda_7$ are indeed repeated, whereas $\lambda_4$ and $\lambda_5$ are convergent eigenvalues despite their temporary clustering.

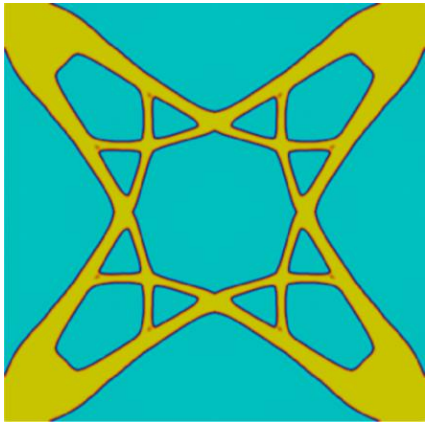
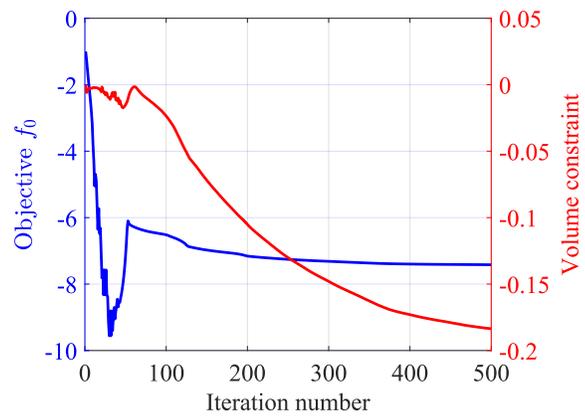

(a) Maximization of $\overline{\Lambda}_1$

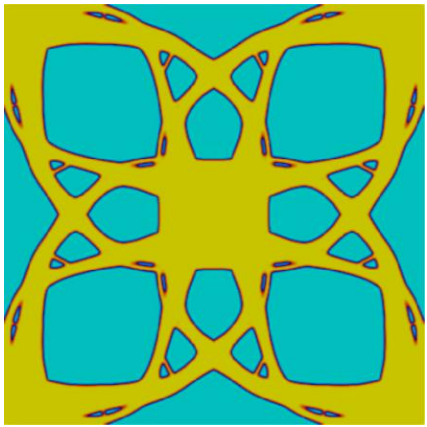
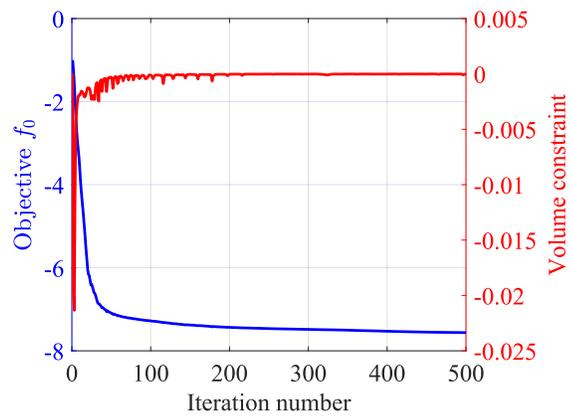

(b) Maximization of $\overline{\Lambda}_2$

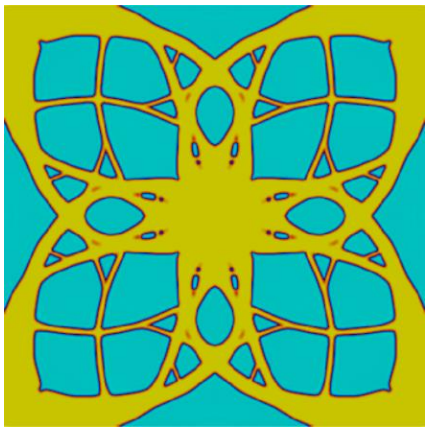
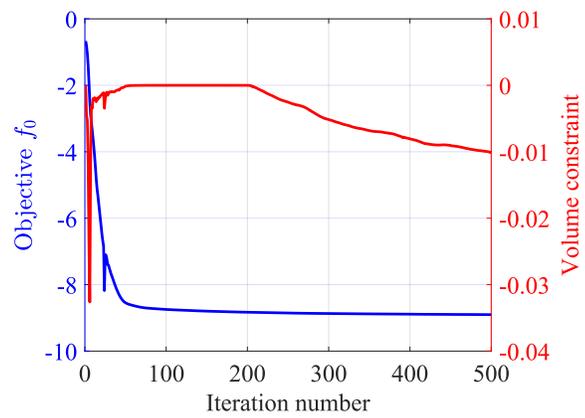

(c) Maximization of $\overline{\Lambda}_2 - \overline{\Lambda}_1$

Figure 17. Square block optimized designs (left) and convergence histories (right) (mesh: 400×400).

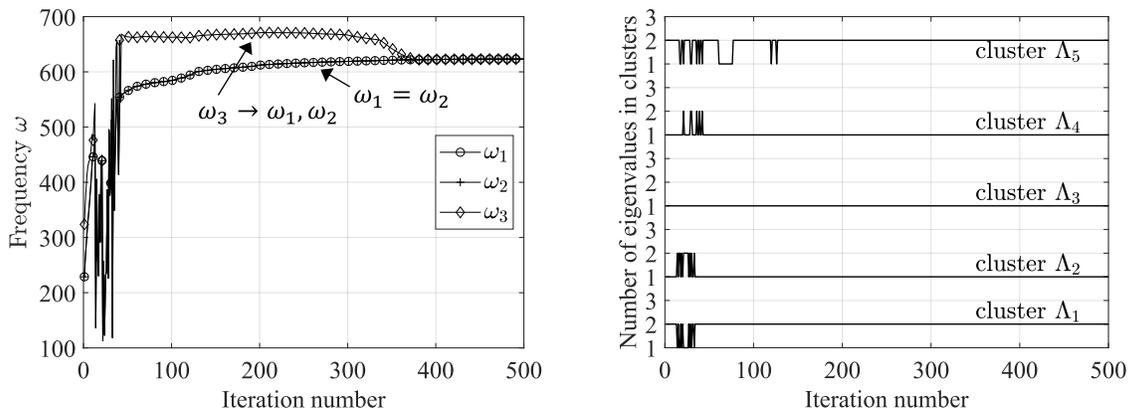

(a) Maximization of $\overline{\Lambda}_1$

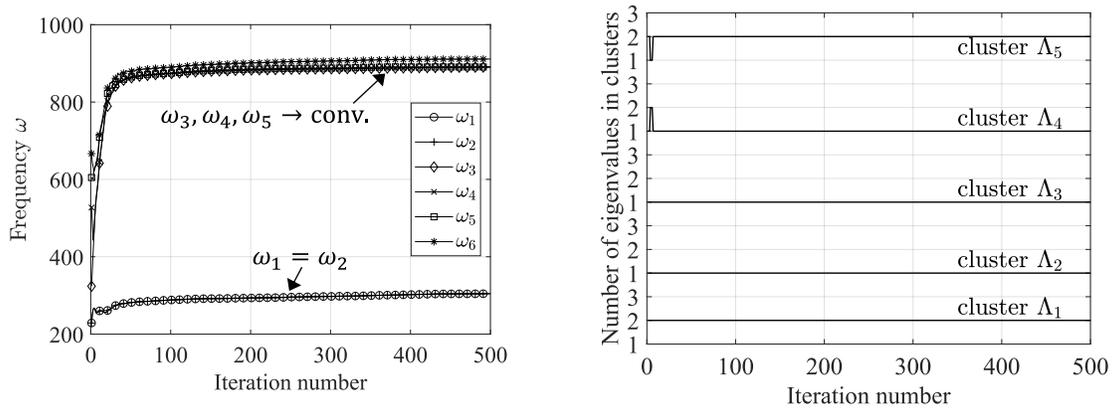

(b) Maximization of $\overline{\Lambda}_2$

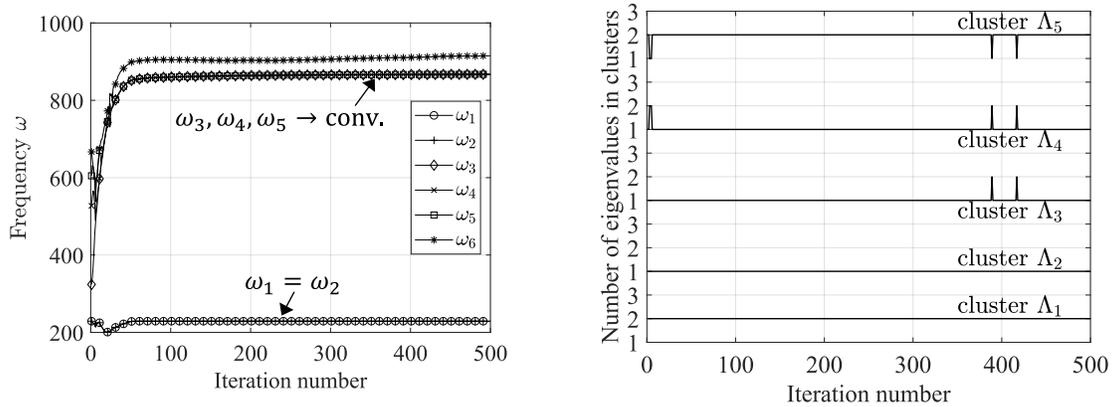

(c) Maximization of $\overline{\Lambda}_2 - \overline{\Lambda}_1$

Figure 18. Square block frequency histories (left) and eigen-cluster histories (right).

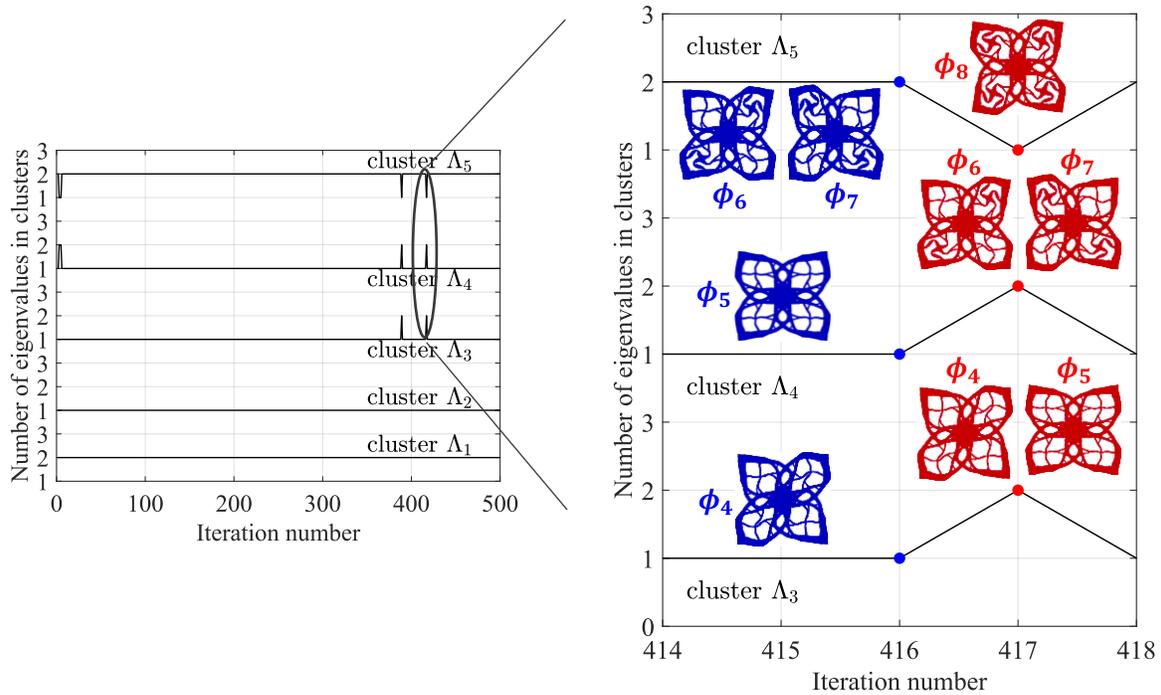

Figure 19. Change in clusters: eigen-cluster history (left) and zoomed-in view (right) with eigenmode plots.

### 3.1.2 Simply supported beam

The presented numerical results have demonstrated the effectiveness of the proposed cluster mean approach in scenarios involving repeated eigenvalues. This proposed method is still valid when eigenvalues remain mostly simple or convergent without repeated clusters. To illustrate this case, the simply supported beam is considered. The design domain is discretized into 800×100 4-node quadrilateral elements (Figure 20(left)). Poisson's ratio is 0.3, the density filter radius is 0.055, and the volume fraction is restricted at 0.5. The Young's modulus ($E_s$) and the mass density ($\rho_s$) of the solid phase are set to 1.0E+7 and 1.0, respectively. Notably, the design space admits ¼-symmetry and such symmetry is enforced. As a comparison study, two different cluster relative tolerances, $\epsilon_{tol} = 1.0\text{E-}8$ and $\epsilon_{tol} = 1.0\text{E-}4$, are used to maximize the fundamental cluster mean $\overline{\Lambda}_1$ of the simply supported beam. This example aims to investigate the effects of tolerance values on the optimization results.

The optimized topologies for two different clustering tolerances are shown in Figure 20(right). By using a small tolerance $\epsilon_{tol}$ =1.0E-8 as the selected default value in this study, the objective function $f_0$ shows smooth convergence (Figure 21(a)). However, the increased tolerance $\epsilon_{tol}$ =1.0E-4, leads to oscillations in the objective function $f_0$ and volume constraint $f_1$ (Figure 21(c)). Furthermore, the eigen-cluster history exhibits continuous oscillations in eigen-cluster $\Lambda_1$ (Figure 21(b)(right)). This is because, with the increased tolerance, the algorithm classifies the two convergent eigenfrequencies, $\omega_1$ and $\omega_2$, into the same cluster in one iteration, and immediately separates them into two clusters in the next iteration. On the other hand, the eigen-clusters with $\epsilon_{tol} = 1.0E-8$ remains simple throughout the optimization. This means that throughout the optimization, all the eigenfrequencies are treated as simple in the sensitivity calculation despite the closeness of eigenfrequencies $\omega_1$ and $\omega_2$, and the simple eigenvalues depend smoothly on the design variables. Moreover, the optimized first eigenvalue with $\epsilon_{tol} = 1.0E-8$ has a slightly higher value $\omega_1^* = 174.4$ than the optimized first eigenvalue $\omega_1^* = 174.0$ with $\epsilon_{tol} = 1.0E-4$. One explanation is that in the steps where $\omega_1$ and $\omega_2$ are treated as repeated eigenvalues, the optimizer attempts to increase the mean of the two. However, the fundamental eigen-cluster $\Lambda_1$ contains only $\omega_1$. Consequently, it is more effective for the optimizer to increase $\omega_1$ instead of the average of $\omega_1$ and $\omega_2$. This comparison study shows that the increased tolerance can classify two convergent simple eigenvalues as repeated and can adversely affect the optimization performance. It is noted that the average of the two simple eigenvalues remains differentiable. Therefore, the result with $\epsilon_{tol} = 1.0E-4$ is still correct. Nonetheless, $\epsilon_{tol} = 1.0E-8$ is selected as the cluster tolerance in this study to avoid any undesirable clustering of convergent eigenvalues.

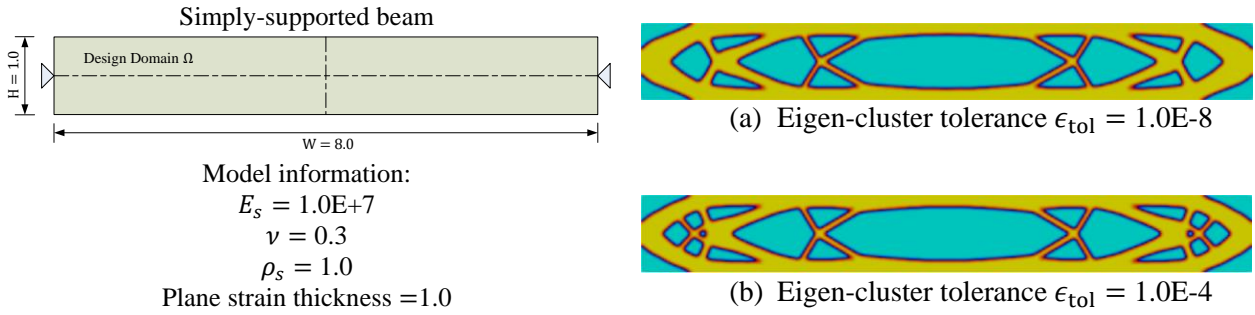

Figure 20. Simply supported beam model (left) and maximization of the fundamental cluster mean $\overline{\Lambda}_1$ results (right) with two different eigen-cluster tolerances (mesh: 800×100).

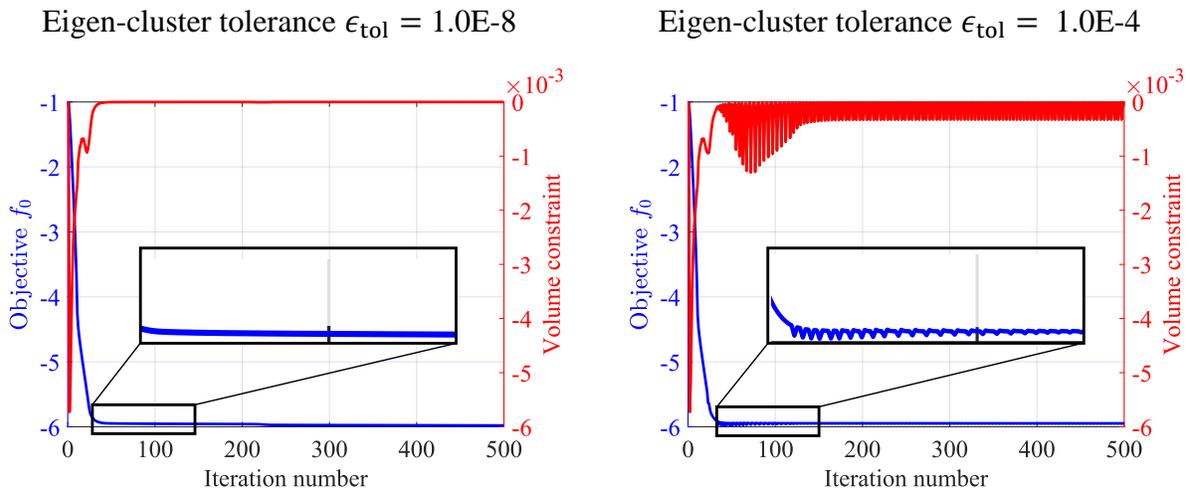

(a) Convergence histories

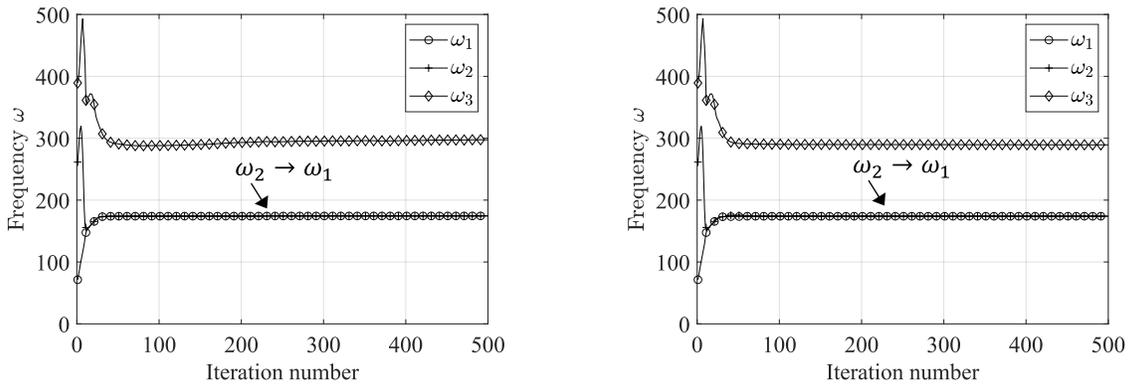

(b) Frequency histories

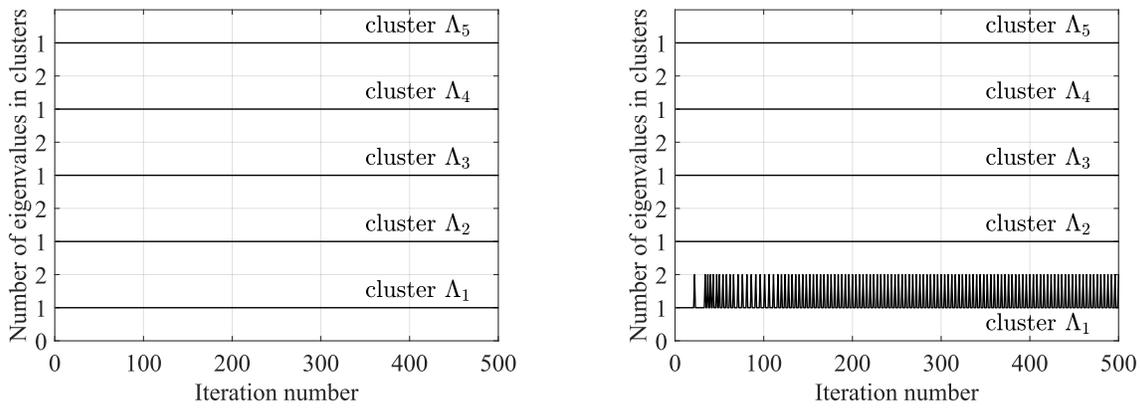

(c) Eigen-cluster histories

Figure 21. Maximization of the fundamental cluster mean $\overline{\Lambda}_1$ with different cluster tolerances.

### 3.1.3 Clamped beam with center point mass

The design domain of the clamped beam (Figure 22) with center point mass is discretized into 800×100 4-node quadrilateral elements. Poisson's ratio is 0.3, the density filter radius is 0.045, and the volume fraction is restricted at 0.45. The design space admits ¼-symmetry, and such symmetry is enforced in the design. The optimized designs of the clamped beam for the first fundamental eigen-cluster mean $\overline{\Lambda}_1$, third eigen-cluster mean $\overline{\Lambda}_3$, and the bandgap $\overline{\Lambda}_3 - \overline{\Lambda}_2$ and their corresponding convergence histories are shown in Figure 22 and Figure 23.

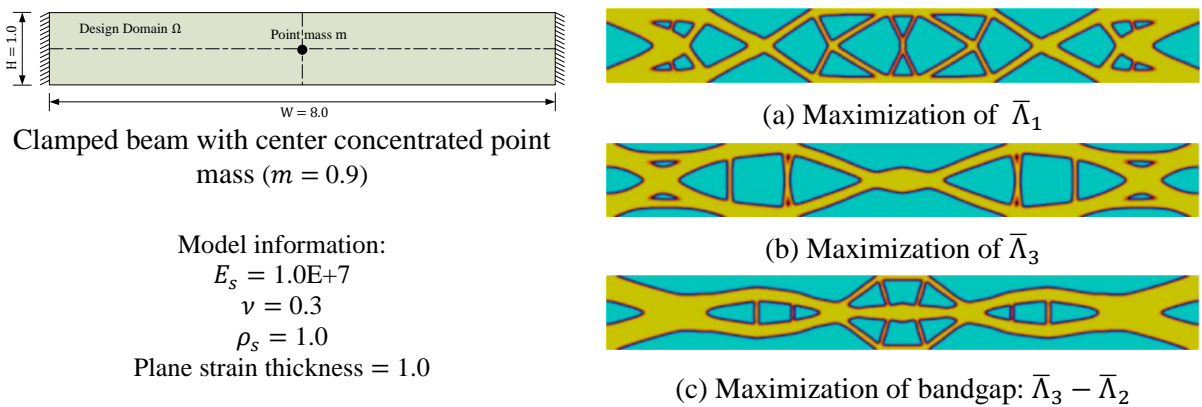

Figure 22. Clamped beam with center point mass (left) and optimized designs (right) (mesh: 800×100).

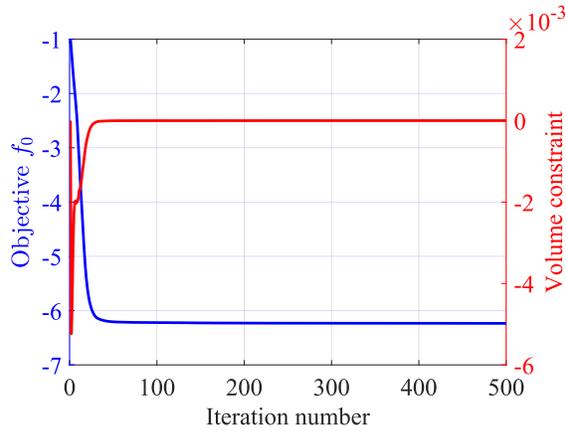
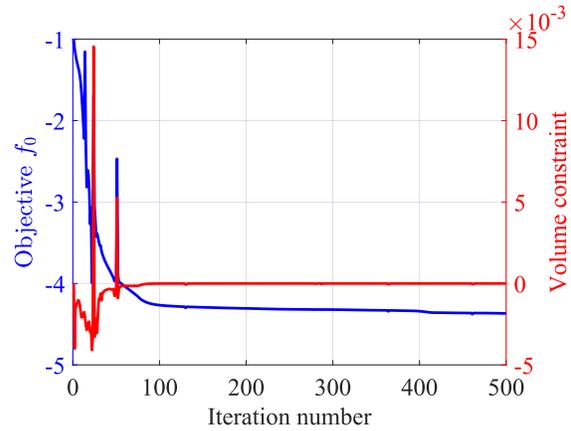

(a) Maximization of $\overline{\Lambda}_1$     (b) Maximization of $\overline{\Lambda}_3$

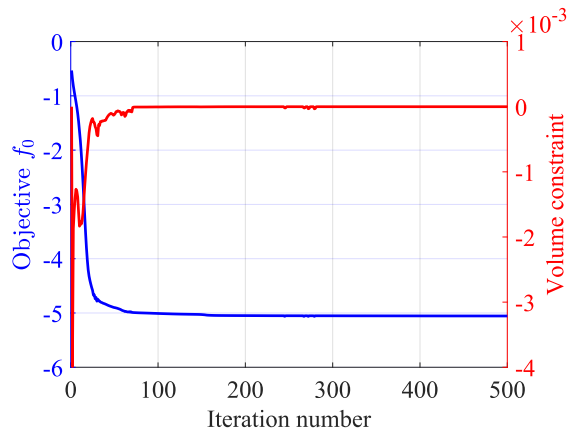

(c) Maximization of $\overline{\Lambda}_3 - \overline{\Lambda}_2$

Figure 23. Convergence histories (clamped beam with center point mass).

Notably, in this case where the third cluster mean $\overline{\Lambda}_3$ is maximized, eigenfrequencies $\omega_3$, $\omega_4$ and $\omega_5$ come close Figure 24(b)(left) but remain separated according to the cluster history shown in Figure 24(b)(right). In the bandgap maximization case, the initially faraway eigenfrequencies $\omega_3$ and $\omega_4$ come close but mostly stay in separate clusters during the optimization Figure 24(c)(right). In the absence of such eigen-cluster information as in the case of using directional derivatives [18] or maximizing a set of eigenvalues without the use of eigen-clusters [49], there is a risk of erroneously asserting multiplicities of 3 and 2 for the eigenvalues in these two cases, solely through visual examination of Figure 24(b)(left) and Figure 24(c)(left), respectively. Again, these eigenvalues are convergent but not repeated.

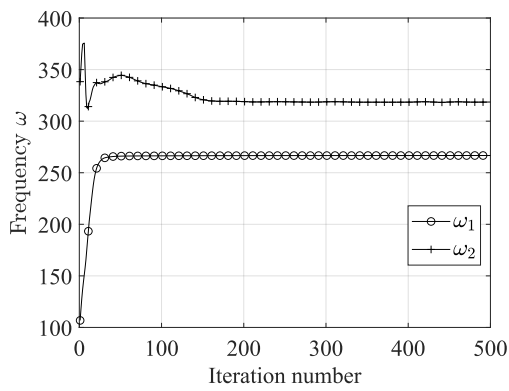 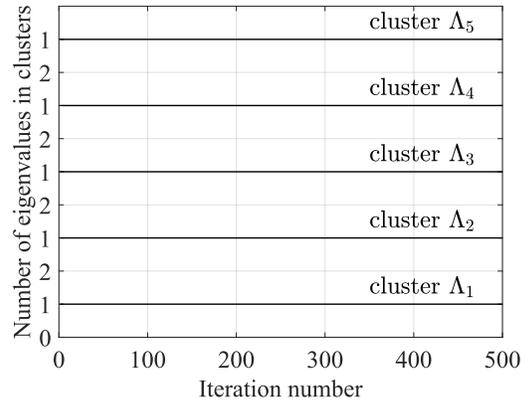

(a) Maximization of $\overline{\Lambda}_1$

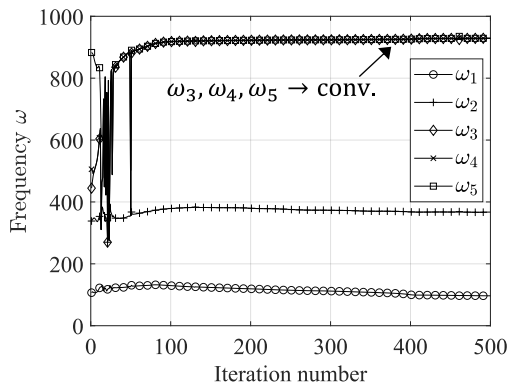 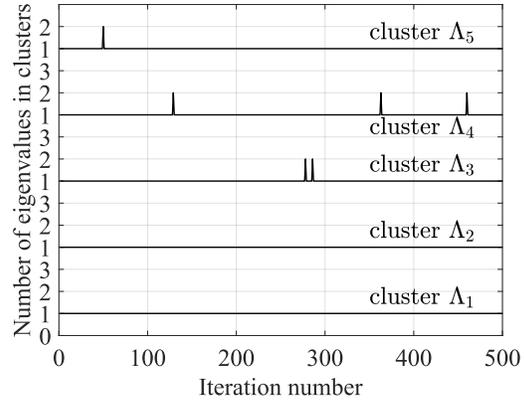

(b) Maximization of $\overline{\Lambda}_3$

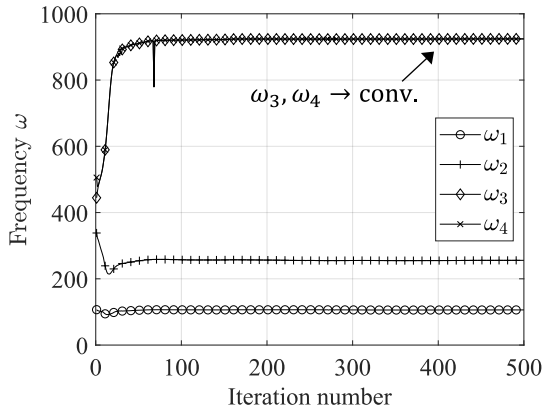 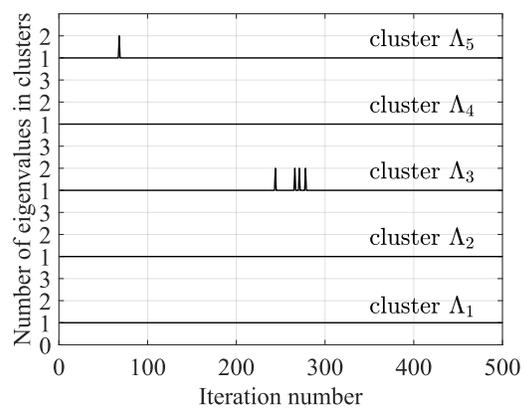

(c) Maximization of $\overline{\Lambda}_3 - \overline{\Lambda}_2$

Figure 24. Clamped beam with center point mass eigenvalue histories (left) and eigen-cluster histories (right).

*3.1.4 Clamped beam with point mass on the lower edge*

The design domain of the clamped beam with point mass on the lower edge is illustrated in Figure 25. The design domain admits ½-symmetry about the vertical centerline, and such symmetry is enforced in the optimization. The model parameters of the example are the same as those in Figure 22(left) except for the location of the point mass. The density filter radius is 0.045, and the volume fraction is restricted at 0.45.

The optimized designs of the clamped beam for the first fundamental eigen-cluster mean $\bar{\Lambda}_1$ and the bandgap $\bar{\Lambda}_3 - \bar{\Lambda}_2$ and their corresponding convergence histories are shown in Figure 26. In this case, when the fundamental cluster mean $\bar{\Lambda}_1$ is maximized, no multiple eigenvalues or convergent eigenvalues are observed. In the case of bandgap optimization, when $(\bar{\Lambda}_3 - \bar{\Lambda}_2)$ is maximized, it is observed that the initially different eigenfrequencies $\omega_3$ and $\omega_4$ approach each other during the optimization. These two eigenfrequencies $\omega_3$ and $\omega_4$, although converging, do not coincide within the same eigen-cluster except for a few steps of temporary numerical clustering.

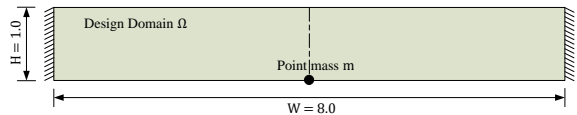

Figure 25. Clamped beam with a concentrated point mass ($m = 0.9$) on the lower edge of the design domain (mesh: 800×100).

Optimized designs

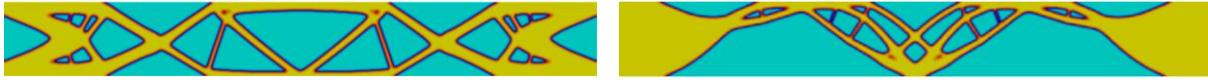

(a) Maximization of $\overline{\Lambda}_1$          (b) Maximization of $\overline{\Lambda}_3 - \overline{\Lambda}_2$

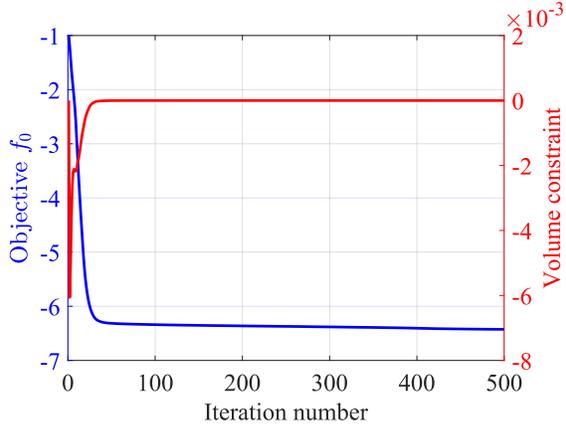 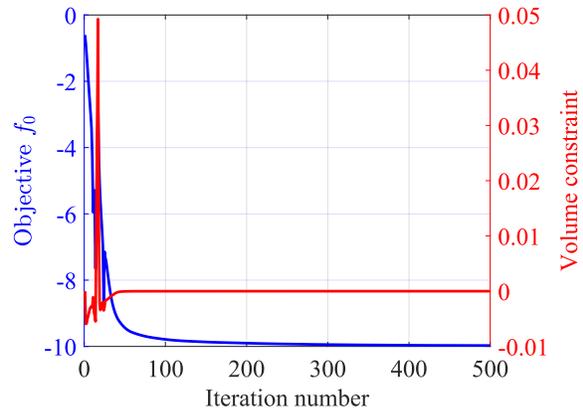

Convergence histories

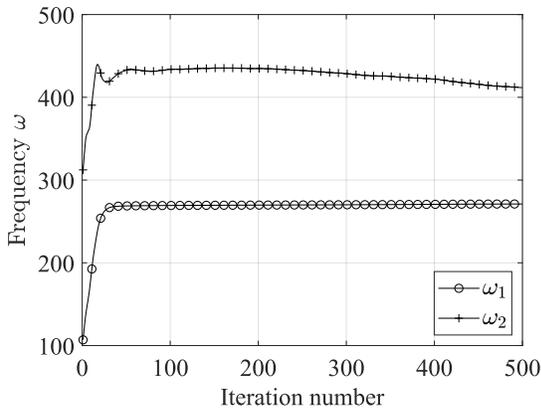 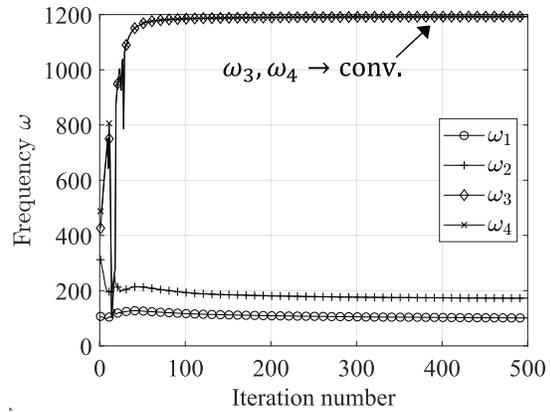

Frequency histories

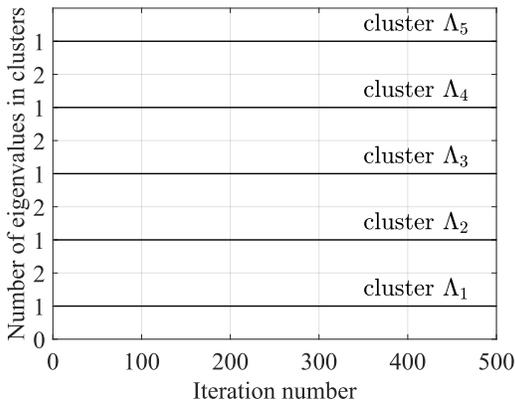 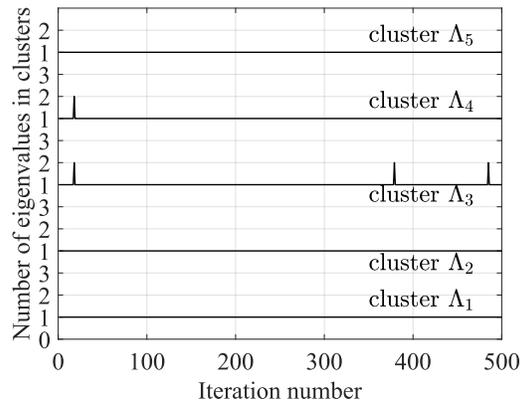

Eigen-cluster histories

Figure 26. Clamped beam with point mass on the lower edge maximization of the fundamental cluster mean $\overline{\Lambda}_1$ and the bandgap $(\overline{\Lambda}_3 - \overline{\Lambda}_2)$.

## 3.2 3D Examples

In this section, the application of the proposed optimization formulation is demonstrated on 3D examples consisting of the maximization of the fundamental eigen-cluster mean and the bandgap. The Young's modulus $E_s$, Poisson's ratio $\nu$ and the mass density $\rho_s$ are taken as 200 GPa, 0.3, and 7800 kg/m$^3$, respectively for all 3D cases. For visualization purposes, a cutoff design value of 0.5 is used to distinguish between solid and void phases in the optimized topology.

### 3.2.1 3D Block with point mass

In this example, the fundamental eigen-cluster mean $\overline{\Lambda}_1$ of the 3D block with a point mass is maximized. The design domain is shown in Figure 27 with the dashed lines indicating the planes of symmetry enforced in the design. The boundary condition includes fixed supports at each corner of the bottom face of the block. The design domain is discretized into 80×80×40 8-node brick elements. The density filter radius is selected as 0.01875 m.

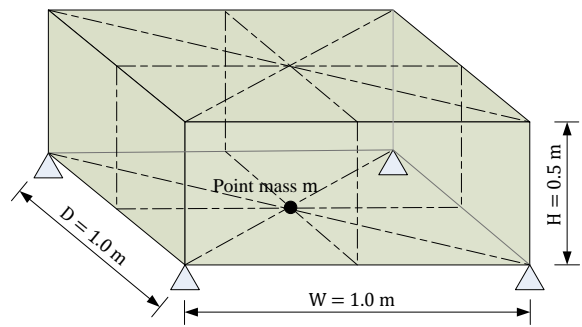

Figure 27. Design domain of the 3D block with point mass ($m = 5000$ kg) in the center of the bottom face (mesh: 80×80×40).

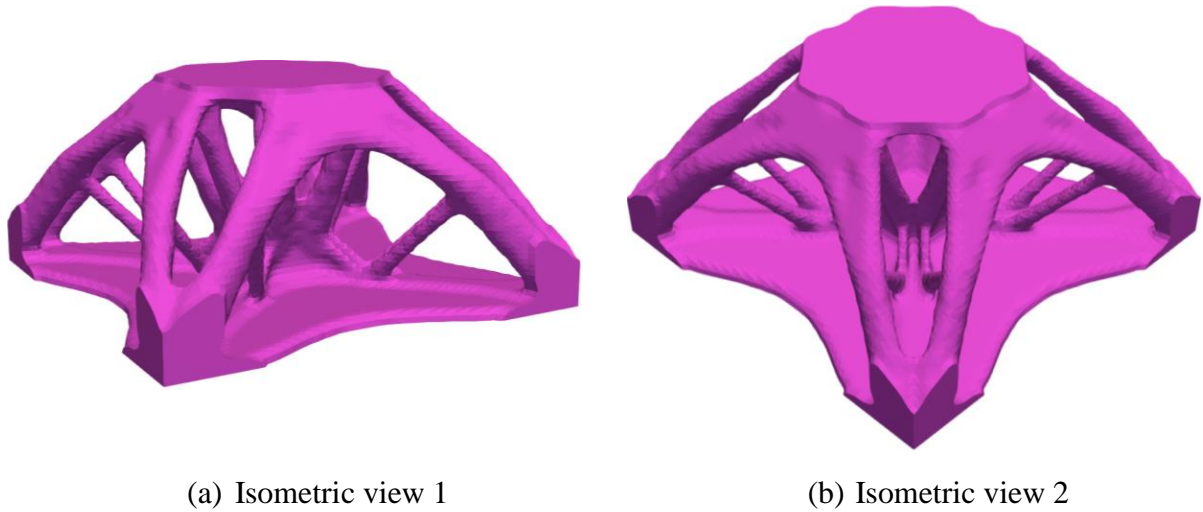

(a) Isometric view 1            (b) Isometric view 2

Figure 28. Maximization of the fundamental cluster mean $\overline{\Lambda}_1$ of the 3D block: optimized designs with two isometric views.

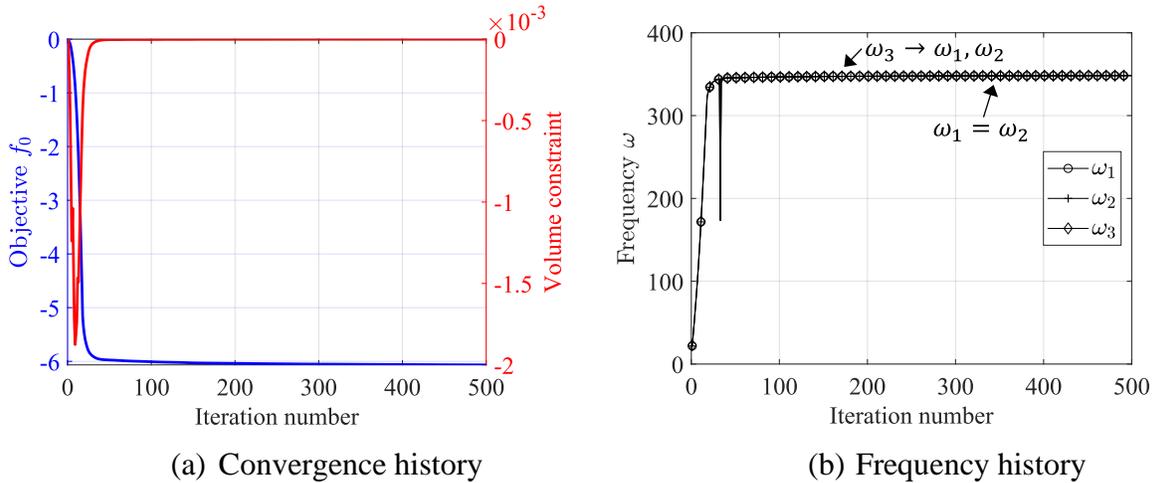

(a) Convergence history           (b) Frequency history

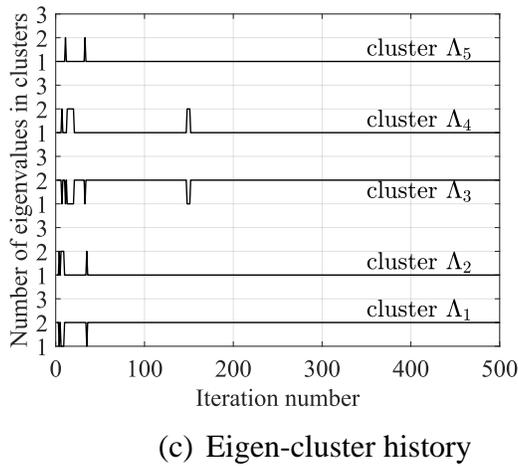

(c) Eigen-cluster history

Figure 29. Maximization of the fundamental cluster mean $\overline{\Lambda}_1$ of the 3D block with point mass.

The two isometric views of the optimized design are shown in Figure 28. The optimization histories for the maximization of the fundamental cluster mean $\bar{\Lambda}_1$ are summarized in Figure 29. The target eigen-cluster $\Lambda_1$ has a multiple eigenvalues (Figure 29(c)) with multiplicity 2, whereas the eigenfrequency $\omega_3$ is a convergent eigenfrequency (Figure 29(b)) for this test case.

*3.2.2 3D Beam with point mass*

In this example, the fundamental eigen-cluster mean $\bar{\Lambda}_1$ of the 3D beam containing a point mass is maximized. Figure 30 depicts the design domain and support conditions, with dashed lines representing the enforced planes of symmetry. The boundary condition consists of fixed supports at every corner of the 3D beam. The design domain is discretized into 50×150×50 8-node brick elements. The density filter radius is selected as 0.03 m.

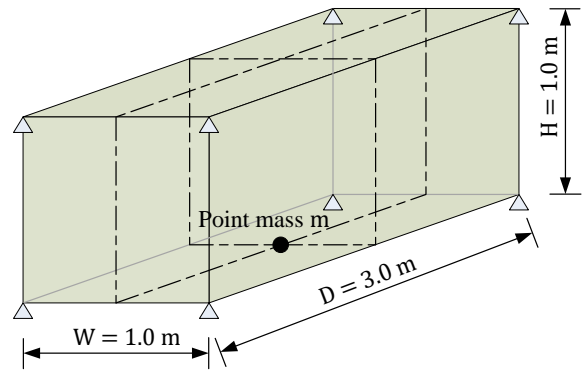

Figure 30. Design domain of the 3D beam with point mass ($m$ = 20,000 kg) in the center of the bottom face (mesh: 50×150×50).

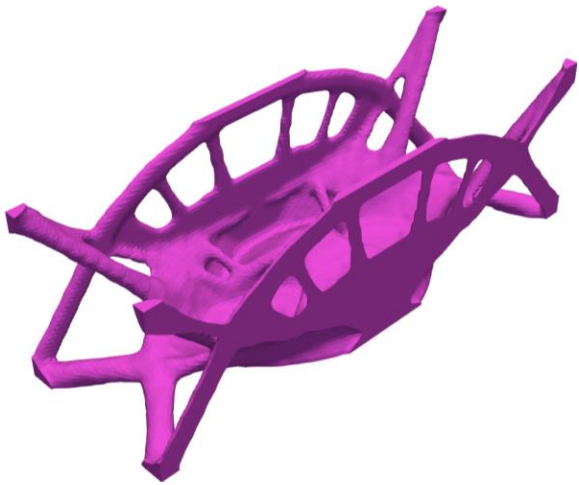
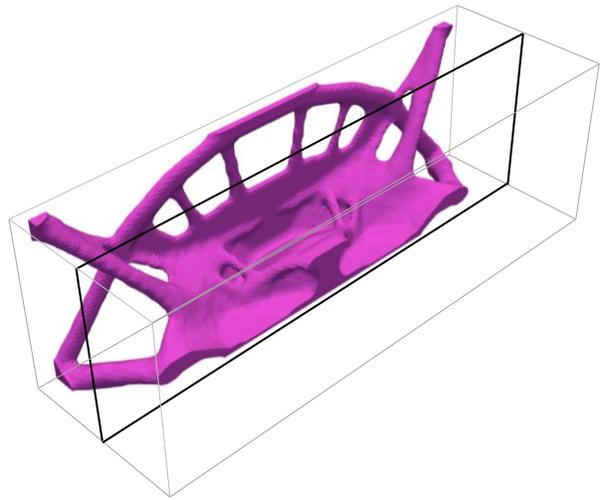

(a) Isometric view  (b) Isometric with cut view

Figure 31. Maximization of the fundamental eigen-cluster mean $\overline{\Lambda}_1$ of the 3D beam with point mass.

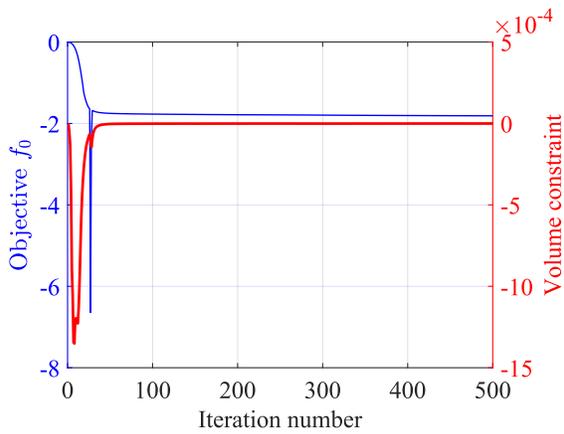
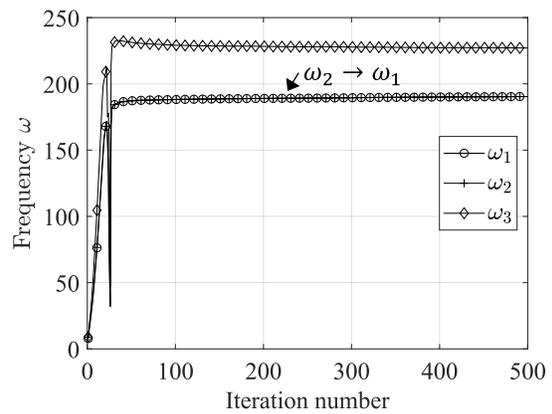

(a) Convergence history  (b) Frequency history

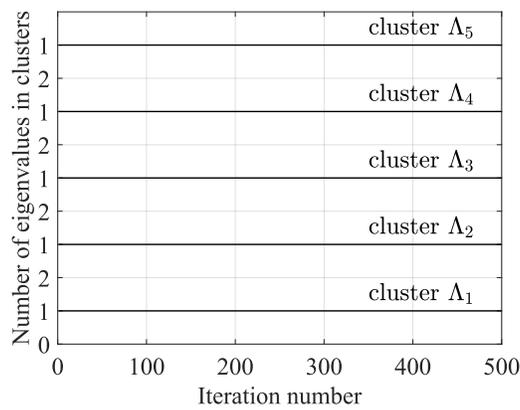

(c) Eigen-cluster history

Figure 32. Maximization of fundamental eigen-cluster mean $\bar{\Lambda}_1$ of the 3D beam with point mass.

Figure 31 shows both an isometric view and a cut view of the optimized design while Figure 32 provides a summary of the optimization histories aimed at maximizing the fundamental cluster mean $\bar{\Lambda}_1$. Notably, no instances of multiple eigenvalues occur during the optimization process. Furthermore, while the eigenfrequency $\omega_2$ converges toward $\omega_1$, the first two eigenfrequencies remain in separate eigen-clusters.

### 3.2.3 3D Clamped beam

In this case, the bandgap $(\bar{\Lambda}_2 - \bar{\Lambda}_1)$ of a 3D clamped beam is maximized. Figure 33 shows the design domain, where dashed lines denote the enforced planes of symmetry. The boundary condition entails clamped supports at the horizontal edges of both ends of the 3D beam. The design domain is discretized into 50×150×50 8-node brick elements. The density filter radius is selected as 0.05 m.

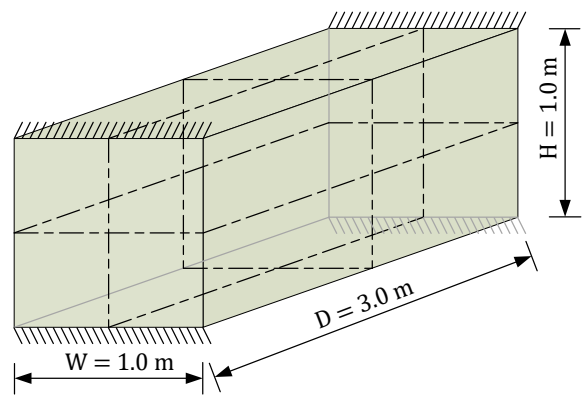

Figure 33. Design domain of the 3D clamped beam (mesh: 50×150×50).

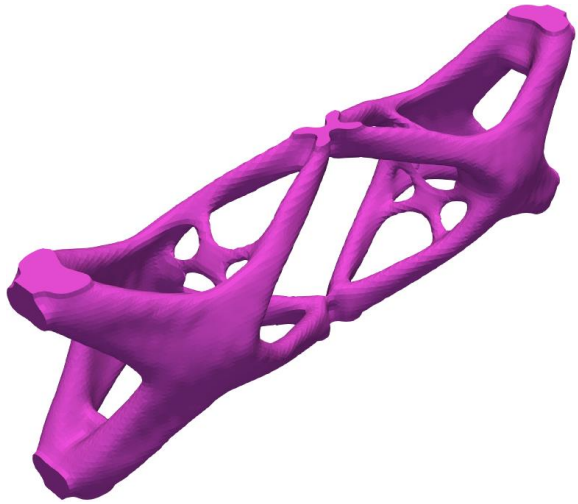 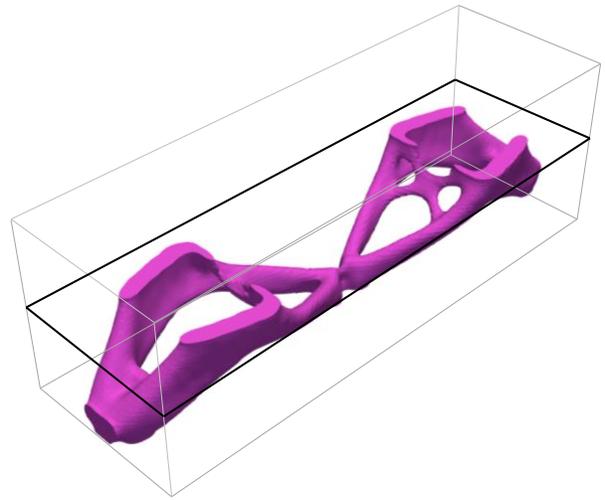

(a) Isometric view  (b) Isometric with cut view

Figure 34. Maximization of the bandgap ($\overline{\Lambda}_2 - \overline{\Lambda}_1$) of the 3D clamped beam.

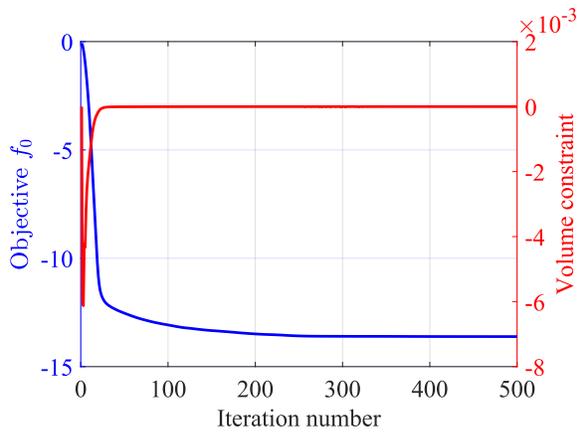 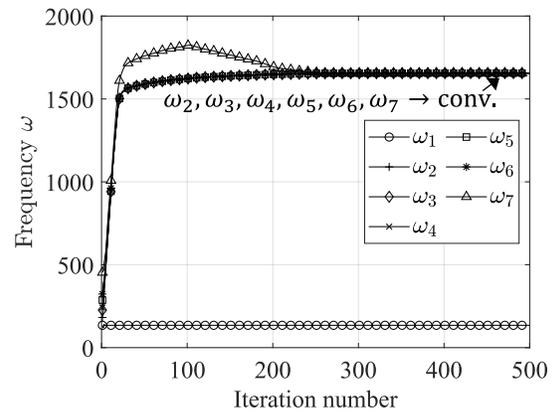

(a) Convergence history  (b) Frequency history

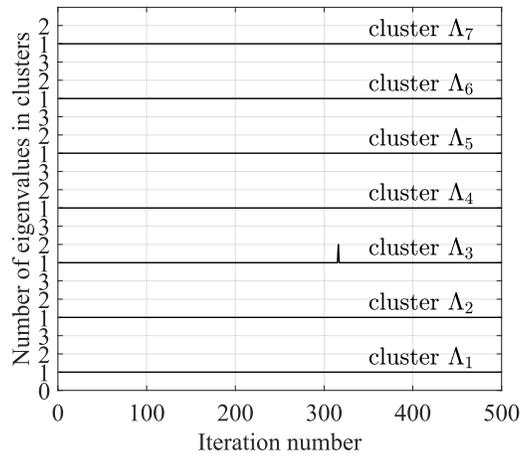

(c) Eigen-cluster history

Figure 35. Maximization of the bandgap ($\bar{\Lambda}_2 - \bar{\Lambda}_1$) of the 3D clamped beam.

Figure 34 shows an isometric view and a cut view of the optimized design while Figure 35 summarizes the optimization histories targeting the maximization of the bandgap ($\bar{\Lambda}_2 - \bar{\Lambda}_1$). Again, there are no occurrences of multiple eigenvalues in the optimized design. Although six eigenfrequencies – $\omega_2$, $\omega_3$, $\omega_4$, $\omega_5$, $\omega_6$ and $\omega_7$ – all converge toward each other; these six eigenfrequencies remain in separate eigen-clusters.

## 3.3  Multi-material Plates

The last set of examples investigates the optimization of multi-material plates. In particular, a square plate ($L = 2.0$ m) is discretized into 400×400 4-node Reissner-Mindlin plate elements (Section 2.1b). The plate thickness is set as 0.01 m. The boundary condition includes pin supports restricting the vertical translation at each corner of the square plate (Figure 36) and ⅛-symmetry is enforced in the design. The density filter radius is 0.01 m. The model parameters vary according to the properties of different selected materials. The representative materials considered in this example include steel, titanium, and aluminum. The Young's modulus and mass density of the materials are summarized in Table 4. The Poisson's ratio for all materials is set to 0.3. A point mass is placed in the center of the square plate.

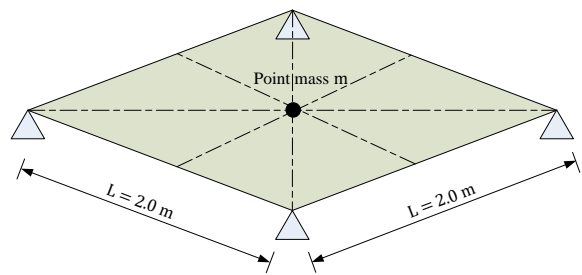

Figure 36. Plate with concentrated point mass in the center design domain (mesh: 400×400).

Table 4. Material properties for multi-material plates.

| Material | Young's modulus (GPa) | Mass density (kg/m$^3$) |
|---|---|---|
| Steel (red) | 210 | 7800 |
| Titanium (green) | 135 | 4500 |
| Aluminum (blue) | 70 | 2700 |

*3.3.1  Bi-material square plate with point mass*

In this example, steel and aluminum are selected to design a bi-material plate. The volume fraction $V_f$ is set as 0.5. The point mass m is set to 21 kg (10% of the mass at the initial step). The optimized designs and the corresponding convergence histories are shown in Figure 37. The objective functions converge smoothly, and the volume fraction constraints are satisfied. In the optimized design figures, the area indicating the material steel is colored red, and the area representing aluminum is colored blue.

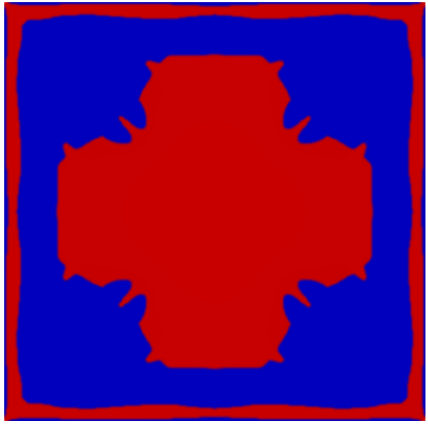 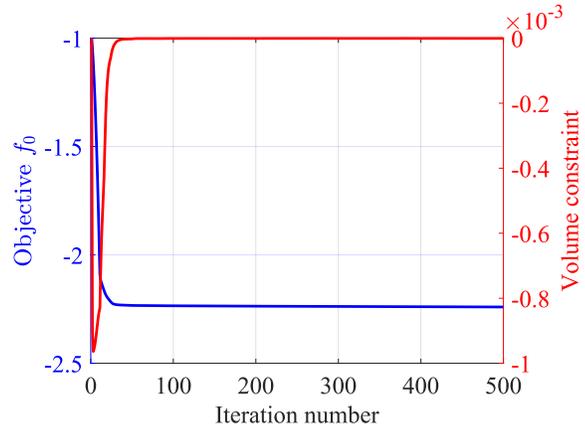

(a) Maximization of $\overline{\Lambda}_1$

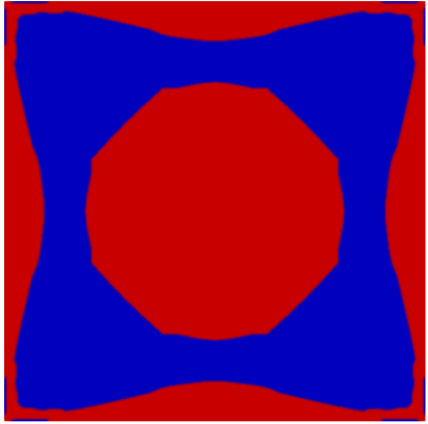 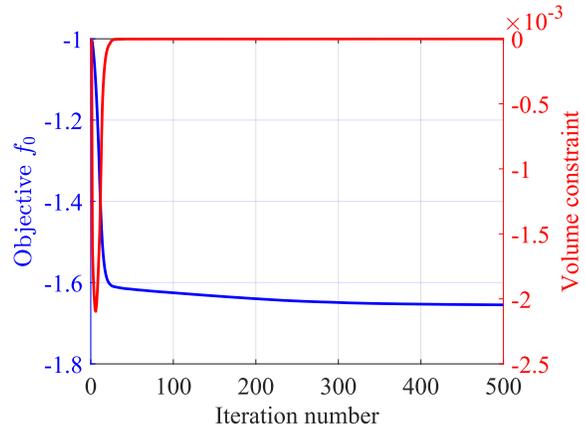

(b) Maximization of $\overline{\Lambda}_2$

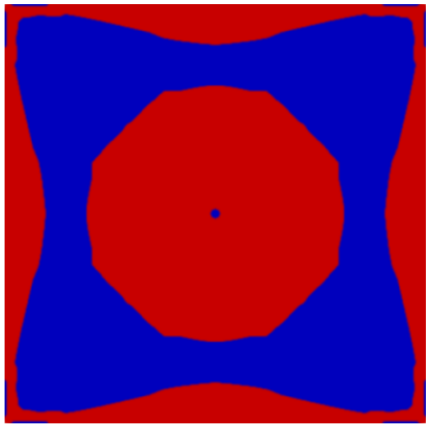 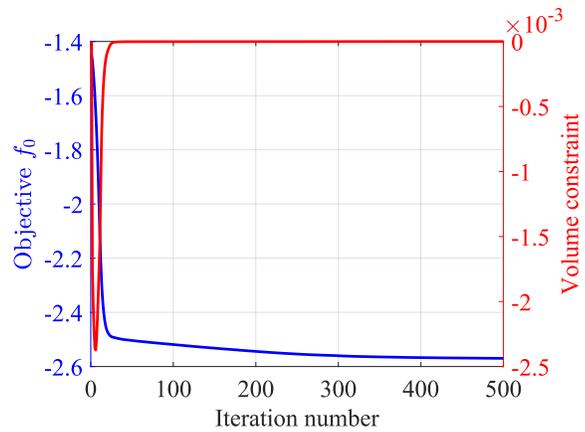

(c) Maximization of $\overline{\Lambda}_2 - \overline{\Lambda}_1$

Figure 37. Bi-material plate with center point mass optimized designs (left) and convergence histories (right).

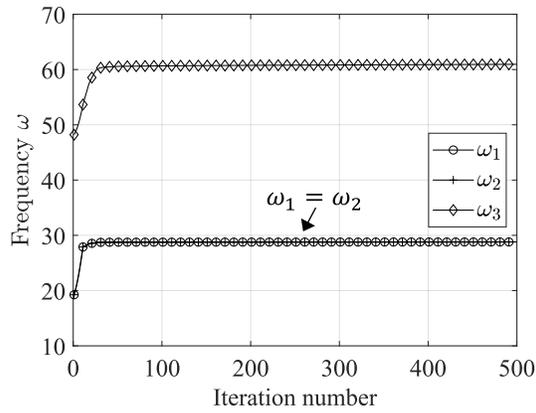 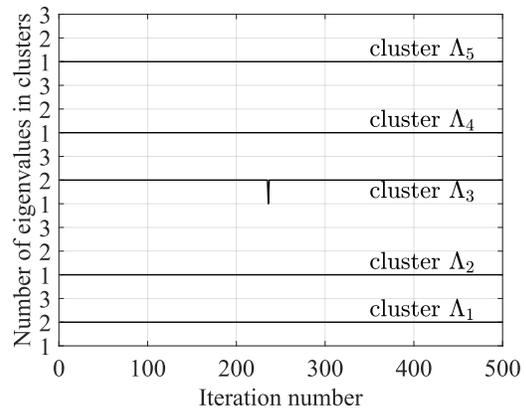

(a) Maximization of $\overline{\Lambda}_1$

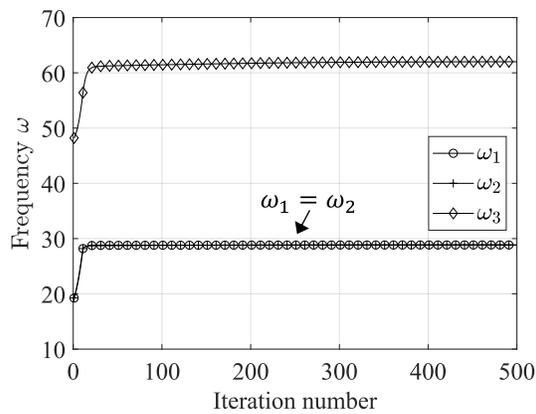 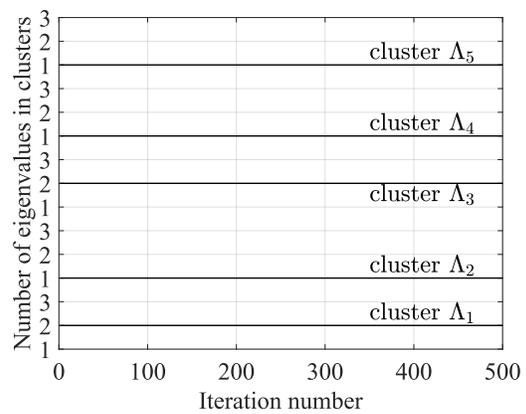

(b) Maximization of $\overline{\Lambda}_2$

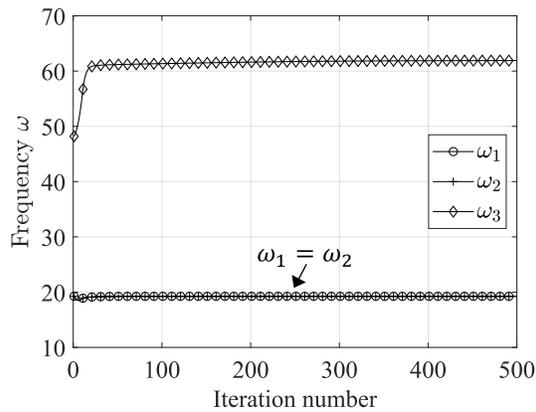 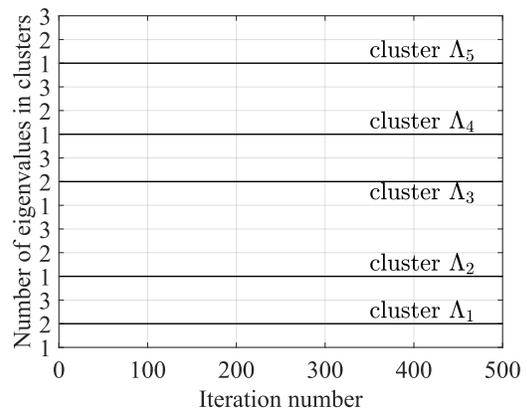

(c) Maximization of $\overline{\Lambda}_2 - \overline{\Lambda}_1$

Figure 38. Bi-material plate with center point mass frequency histories (left) and eigen-cluster histories (right).

Based on Figure 38, the repeated eigenvalues exist in the initial design and remain clustered during the optimization. Additionally, no convergent eigenvalues occur during the optimization.

### 3.3.2 Multi-material square plate with point mass

In this example, steel (red) and titanium (green) are selected for the design of the bi-material + void phase plate, and the corresponding volume constraint thresholds are set as $V_1^f = 0.5$, $V_2^f = 0.25$, respectively. The point mass $m$ is set as 61.5 kg (50% of the mass at the initial step) in the case of the bi-material + void phase. Steel (red), titanium (green), and aluminum (blue) are selected for the design of the tri-material + void phase plate, and the volume fractions are set as $V_1^f = 0.5$, $V_2^f = 0.25$ and $V_3^f = 0.125$, respectively. The point mass $m$ is set as 44.25 kg (50% of the mass at the initial step) in the case of the tri-material + void phase. The optimized designs and convergence histories of the multi-material plate with void phase for the maximization of fundamental eigen-cluster mean $\bar{\Lambda}_1$ are shown in Figure 39. The convergence histories are smooth, and all volume fraction constraints are satisfied in optimized designs. The volume fractions of the three phases of material are denoted as $V_r^f$ (red: steel), $V_g^f$ (green: titanium) and $V_b^f$ (blue: aluminum). The volume fractions of the two material phases are obtained as

$$\begin{aligned} V_r^f &= V_2^f + \bar{f}_{m+2} \\ V_g^f &= V_1^f - V_r^f + \bar{f}_{m+1} \end{aligned} \quad (25)$$

where $\bar{f}_{m+1}$ and $\bar{f}_{m+2}$ are function values of the volume constraints in Eq. (16). Similarly, the volume fractions of the three material phases are given by

$$\begin{aligned} V_r^f &= V_3^f + \bar{f}_{m+3} \\ V_g^f &= V_2^f - V_r^f + \bar{f}_{m+2} \\ V_b^f &= V_1^f - V_r^f - V_g^f + \bar{f}_{m+1} \end{aligned} \quad (26)$$

where $\bar{f}_{m+1}$, $\bar{f}_{m+2}$ and $\bar{f}_{m+3}$ are function values of the volume constraints in Eq. (17).

It can be observed from the eigenfrequency and eigen-cluster histories shown in Figure 40 that the fundamental eigen-cluster $\Lambda_1$ contains two repeated eigenvalues. The two repeated eigenvalues, $\omega_1$ and $\omega_2$, remain clustered during the optimization when the first cluster mean, $\overline{\Lambda}_1$, is maximized.

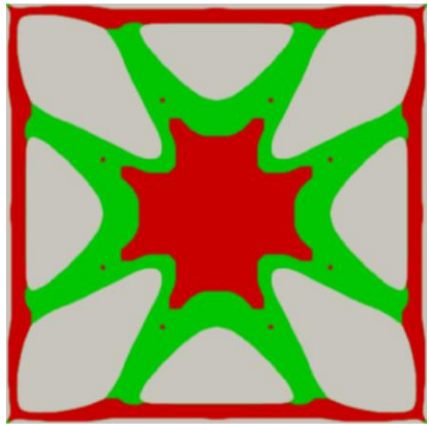
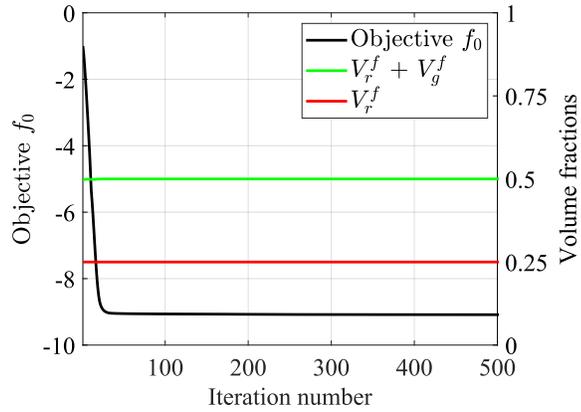

(a) Bi-material plate with void phase

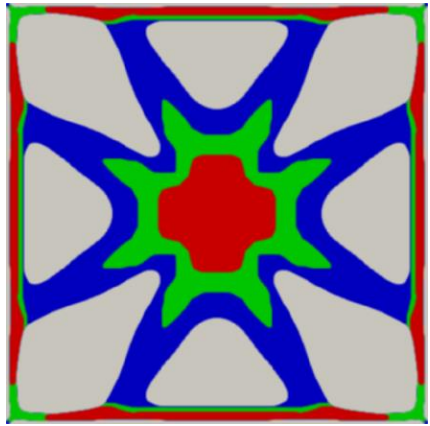
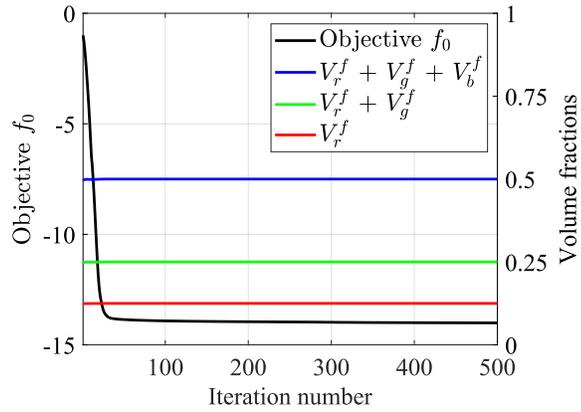

(b) Tri-material plate with void phase

Figure 39. Maximization of the fundamental cluster mean $\overline{\Lambda}_1$ of multi-material plates with void phase: optimized designs (left) and convergence histories (right).

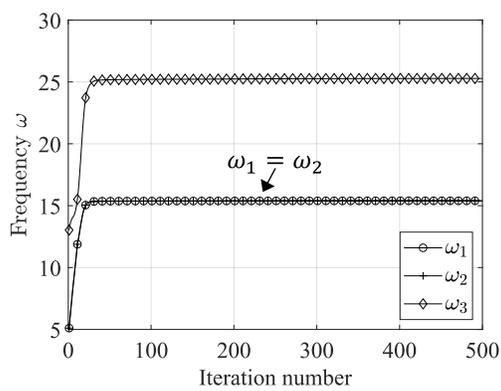
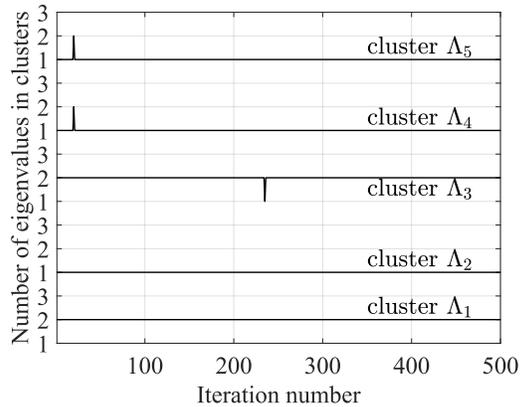

(a) Bi-material plate with void phase

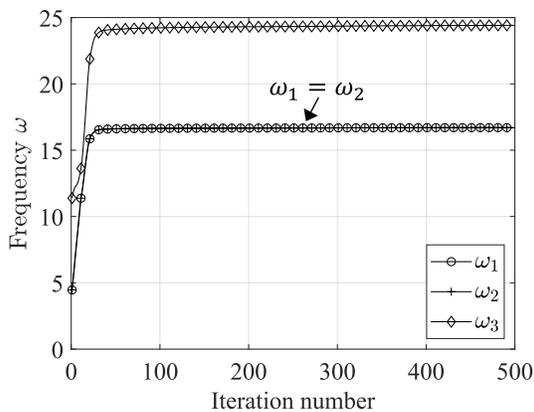
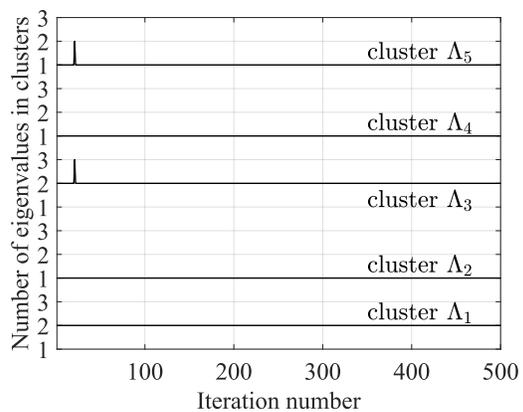

(b) Tri-material plate with void phase

Figure 40. Maximization of the fundamental cluster mean $\overline{\Lambda}_1$ of multi-material plates with void phase: frequency histories (left) and eigen-cluster histories (right).

The results of the maximization of the second eigen-cluster mean $\overline{\Lambda}_2$ of the multi-material plate with void phase are shown in Figure 41. The convergence histories are smooth, and all volume fraction constraints are satisfied. It is observed from Figure 42 that the target eigen-cluster $\Lambda_2$ contains a simple eigenfrequency $\omega_3$, the fundamental eigen-cluster $\Lambda_1$ has repeated eigenfrequency $\omega_1$ and $\omega_2$, and the optimization does not lead to convergent eigenvalues in this case.

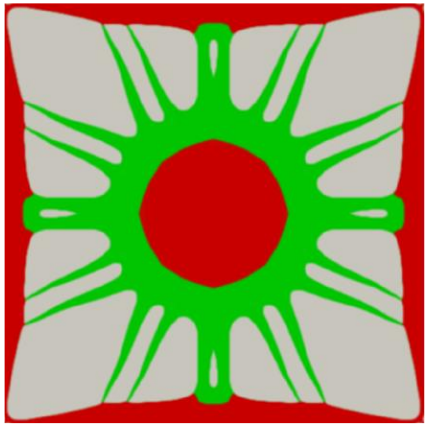
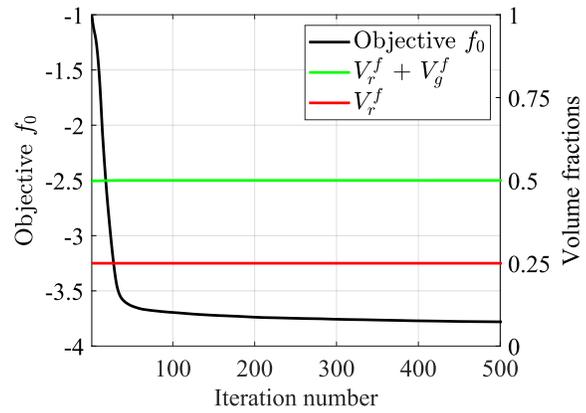

(a) Bi-material plate with void phase

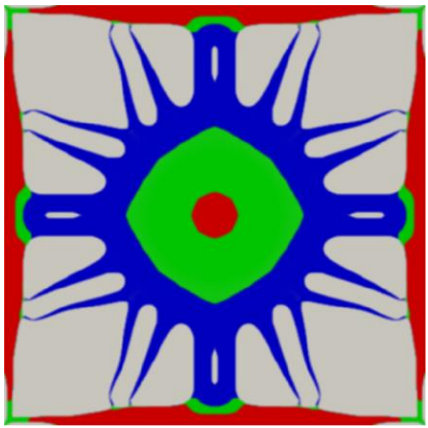
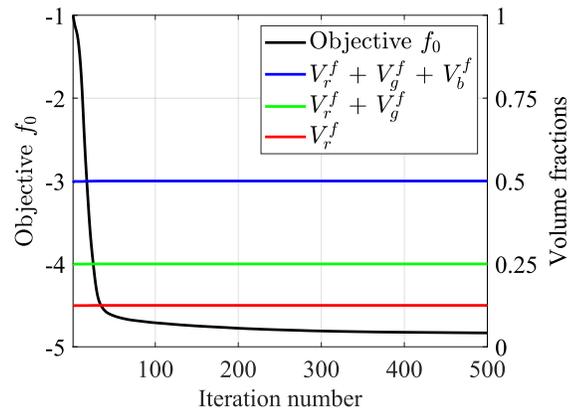

(b) Tri-material plate with void phase

Figure 41. Maximization of the second cluster mean $\overline{\Lambda}_2$ of multi-material plates with void phase: optimized designs (left) and convergence histories (right).

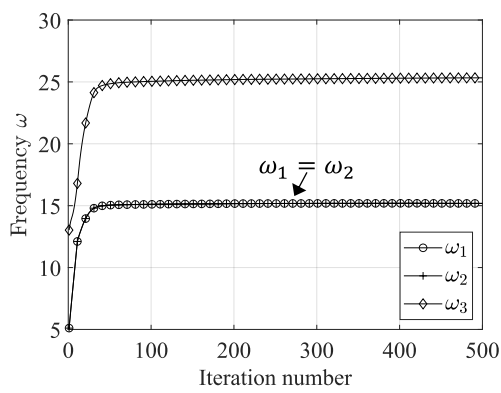 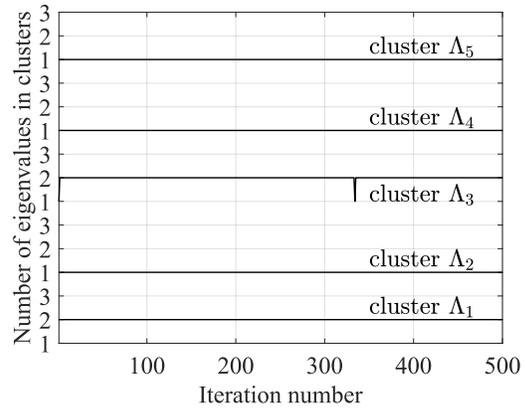

(a) Bi-material plate with void phase

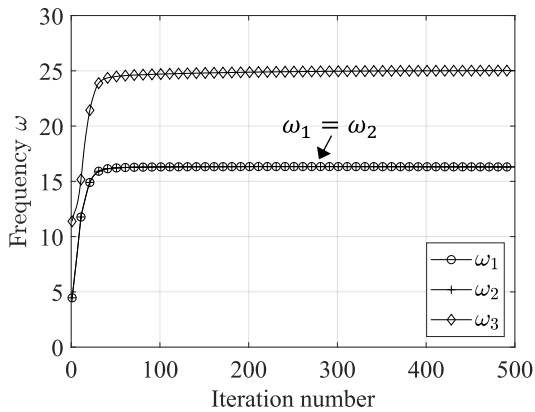 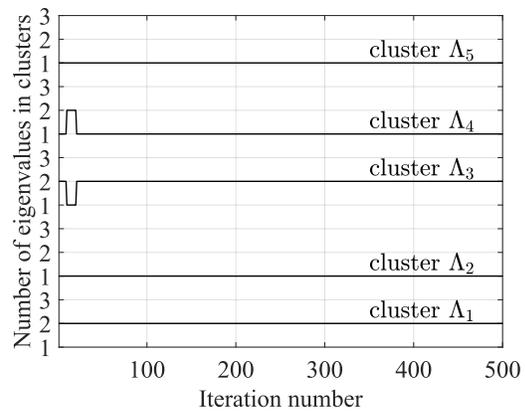

(b) Tri-material plate with void phase

Figure 42. Maximization of the second cluster mean $\overline{\Lambda}_2$ of multi-material plates with void phase: frequency histories (left) and eigen-cluster histories (right).

## 4 Conclusions

In this study, a cluster mean approach is applied to resolve the non-differentiability of multiple eigenvalues for optimizing eigenfrequencies and bandgaps. The cluster mean approach relies on the construction of symmetric functions of repeated eigenvalues. It is shown that a symmetric function of multiple eigenvalues, such as the cluster mean, $p$-norm, and KS functions are *differentiable* only when all the repeated eigenvalues are included in these functions, i.e., the clusters are complete. As these functions are frequently used in eigenvalue topology optimization, care should be taken when employing such symmetric functions to ensure differentiability. This result also resolves the issue in past studies, where different claims regarding the differentiability of such functions were made [28].

Sensitivity results also show that the multiple eigenvalues are differentiable w.r.t the symmetric design variables under certain symmetry conditions. However, this result is problem-dependent, and it may be not easy to predetermine the conditions under which the multiple eigenvalues are differentiable [48]. Notably, the sensitivity results in Section 2.4 confirm that regardless of the enforced symmetry, the cluster mean approach guarantees the differentiability of multiple eigenvalues with respect to both *symmetric* and *all* design variables, providing a reliable method in eigenfrequency topology optimization. Therefore, it is expedient to use cluster means to ensure the differentiability of repeated eigenvalues, as this method has little computational overhead.

The cluster mean approach is combined with the bound variable method to formulate optimization schemes for maximizing eigenfrequencies and bandgaps. The efficacy of the proposed optimization schemes is demonstrated through numerical examples consisting of 2D and 3D solid structures and multi-material plate structures with various enforced symmetries. In *all* instances, smooth convergence histories are observed, and discrete optimized topologies are obtained. It is shown that in many cases, the eigenfrequencies are convergent but remain simple. In some past studies, these cases are sometimes misidentified as repeated eigenvalue cases. Thus, the clustering strategy discussed in this work also serves as a valuable tool for identifying multiple and convergent eigenvalues during optimization, clarifying ambiguities in the previous literature.

**Acknowledgments**

The presented work is supported in part by the U.S. Department of Energy by Lawrence Livermore Laboratory under Contract #B652057. Any opinions, findings, conclusions, and recommendations expressed in this article are those of the authors and do not necessarily reflect the views of the sponsors.

**Data Availability**

Data sharing does not apply to this article as no datasets were generated or analyzed during the current study.